\documentclass[10pt, aps, prl, longbibliography, twocolumn, superscriptaddress, showpacs, floatfix]{revtex4-1}

\usepackage{here}
\usepackage{amsmath}
\usepackage{graphicx}
\usepackage{braket}
\usepackage{lettrine}
\usepackage{color}
\usepackage{times}
\usepackage{mathtools}
\usepackage[T1]{fontenc}

\begin{document}

\setlength{\textfloatsep}{15pt}
\setlength{\parskip}{0pt}

\title{
Orbital Torque: Torque Generation by Orbital Current Injection
}

\author{Dongwook Go}
\affiliation{Department of Physics, Pohang University of Science and Technology, Pohang 37673, Korea}

\affiliation{Basic Science Research Institute, Pohang University of Science and Technology, Pohang 37673, Korea}

\affiliation{Peter Grünberg Institut and Institute for Advanced Simulation, Forschungszentrum Jülich and JARA, 52425 Jülich, Germany}

\author{Hyun-Woo Lee}
\email{hwl@postech.ac.kr}
\affiliation{Department of Physics, Pohang University of Science and Technology, Pohang 37673, Korea}

\begin{abstract}

We propose a mechanism of torque generation by injection of an orbital current, which we call {\it orbital torque}. In a magnetic bilayer consisting of a nonmagnet (NM) and a ferromagnet (FM), we consider a situation where the spin-orbit coupling (SOC) is present only in the FM. Although the SOC is absent in the NM, the orbital Hall effect can arise in the NM. When the resulting orbital Hall current is injected to the FM, the SOC of the FM converts the orbital angular momentum into spin, which exerts torque to the magnetization of the FM. Remarkably, even for small SOC strength comparable to that of $3d$ FMs, the orbital torque can be comparable to the spin torque induced by the spin Hall effect of the NM with strong SOC. This provides a way to experimentally probe the OHE and opens a new venue to achieving spin-torque devices based on light elements that exhibit gigantic orbital response. Experimental implications are discussed.
\end{abstract}

%\date{\today}
\maketitle	

Spin injection into a ferromagnet (FM) generates a spin torque (ST) on magnetic moments of the FM by the angular momentum transfer from the spin of injected conduction electrons. For ST generation, a spin current source is needed. A popular source is a nonmagnet (NM) with strong spin-orbit coupling (SOC), which exhibits sizable spin Hall effect (SHE). The ST of the SOC origin is called spin-orbit torque~\cite{Pi2010, Miron2010, Miron2011a, Miron2011b, Pai2012, Liu2012a, Liu2012b, Kim2012, Haney2013a, Haney2013b, Garello2013, Garello2014, Kurebayashi2014, Hayashi2014, Yu2014, Freimuth2014, Kim2017, Mahfouzi2018}, which has drawn considerable attention as a powerful means to electrically control magnetic configurations. 

%such as current-induced magnetization switching~\cite{Miron2011b, Liu2012a, Liu2012b, Yu2014, Garello2014} and domain wall motion~\cite{Miron2011a, Ryu2013, Emori2013, Martinez2013}.

%Similar to the SHE, the orbital Hall effect (OHE) \cite{Bernevig2006, Tanaka2008, Kontani2009, Go2018, Jo2018} is also possible, which describes electrical generation of a transverse orbital current. 

Similar to the SHE, the orbital Hall effect (OHE) allows for electrical generation of a transverse orbital current. In transition metals, for example, electron wavefunctions near atomic cores have mainly $d$ character, and superpositions such as $d_{zx}\pm i d_{yz}$ carry the orbital angular momentum $L_z=\pm \hbar$. A flow of wavepackets with such superposed wavefunctions generates an orbital current. Considering that an orbital current carries the angular momentum just like a spin current does, it is reasonable to expect that injection of an orbital current (or orbital injection in short) into a FM may generate a torque on local magnetic moments of the FM. We call such torque as {\it orbital torque} (OT), which provides an experimental way to detect the OHE.
Although the OHE has not yet been experimentally verified, theoretical calculations~\cite{Tanaka2008, Kontani2009} on $4d$ and $5d$ transition metals indicate that the orbital Hall conductivities (OHCs) of these NM's are about an order of magnitude larger than the spin Hall conductivities (SHCs). Moreover, our recent theoretical analysis finds that the OHC can be gigantic $\sigma_\mathrm{OH}\sim 10^4 (\hbar/2|e|)(\Omega\cdot{\rm cm})^{-1}$ even in materials with negligible SOC~\cite{Go2018, Jo2018}. Thus the OT also provides a new venue to achieving high torque efficiency in spintronic devices.

%Moreover our recent theoretical analysis~\cite{Go2018} finds the SHE in centrosymmetric systems arises from the combined action of the OHE and the SOC. This motivates experimental attempts to verify the OHE. The OT provides an experimental way to detect the OHE. The OHE may be valuable in view of device applications that demand highly efficient spin/orbital current sources. Since the emergence of the OHE does not require the SOC~\cite{Bernevig2006,Go2018}, not only NM's with strong SOC~\cite{Tanaka2008,Kontani2009} but also NM's with weak SOC can also have gigantic OHCs $\sigma_\mathrm{OH}\sim 10^4 (\hbar/2|e|)(\Omega\cdot{\rm cm})^{-1}$~\cite{Jo2018}. Thus the OHE significantly widens the material choice for NM's.

\begin{figure}[t!]
\includegraphics[angle=0, width=0.35\textwidth]{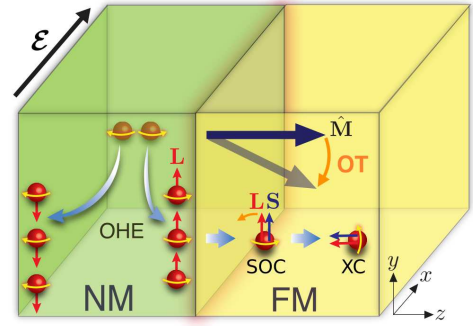}
\caption{
\label{fig:schematics}
Schematic illustration of the OT in a NM/FM bilayer. The orbital Hall current generated in the NM flows into the FM. The SOC of the FM then converts the orbital angular momentum to the spin, which exerts torque to the magnetization $\hat{\mathbf{M}}$.}
\end{figure}

In this Letter, we investigate the theoretical idea of the OT for a NM/FM bilayer structure (Fig.~\ref{fig:schematics}). When an in-plane electric field $\boldsymbol{\mathcal{E}}$ is applied, both OHE and SHE arise in the NM in general \cite{Go2018, Kontani2009, Tanaka2008, Jo2018}. In order to focus on the OT due to the orbital injection, we suppress the SHE by setting the SOC of the NM zero. Then only OHE is induced and a resulting torque in the FM can be identified unambiguously as the OT. We find that the OT indeed arises as long as the SOC of the FM is finite.

%========================================================
%========================================================

\begin{figure}[t!]
\includegraphics[angle=0, width=0.45\textwidth]{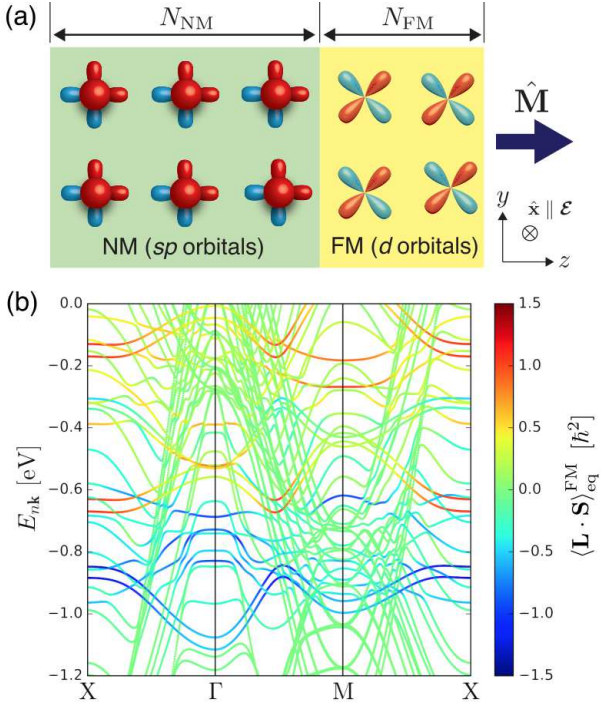}
\caption{
\label{fig:model}
(a) Schematic illustration of the tight-binding model of the NM/FM bilayer. (b) The band structure of the bilayer for $N_{\rm NM}=8$ and $N_{\rm FM}=2$. The color represents
the equilibrium expectation value of the spin-orbit correlation in the FM region $\left\langle \mathbf{L}\cdot\mathbf{S}\right\rangle_{\rm eq}^{\rm FM}$ for each state.
}
\end{figure}

For a quantitative evaluation of the OT, we adopt the tight-binding description of the bilayer with $N_\mathrm{NM}(N_\mathrm{FM})$ atomic layer thick NM(FM) [Fig.~\ref{fig:model}(a)]. We assume both NM and FM to have the simple cubic structure. For the NM, we adopt the $sp$ model that has been used previously~\cite{Go2018} to illustrate the OHE without the SOC. 
In this model, each lattice site can host $s$, $p_x$, $p_y$, and $p_z$ orbitals, and the orbital hydridization, which is crucial for the emergence of the OHE~\cite{Go2018}, arises from the symmetry-allowed nearest neighbor hoppings between $s$ and $p_{x,y,z}$ orbitals.
For the FM, we adopt a trivial $d$ model;  each lattice site can host $d_{xy}$, $d_{yz}$, $d_{zx}$, $d_{z^2}$, and $d_{x^2-y^2}$ orbitals with nearest neighboring hoppings allowed. This $d$ model does not allow any orbital hybridization \cite{Cubic_d_comment} and thus there is no OHE \cite{Go2018, Jo2018}. The $d$ model is augmented by adding the SOC
\begin{eqnarray}
\label{eq:SOC_FM}
H_\mathrm{so}^\mathrm{FM}
=
\frac{\alpha_\mathrm{so}^\mathrm{FM}}{\hbar^2} \mathbf{L} \cdot \mathbf{S},
\end{eqnarray}
and the exchange coupling $H_{\rm xc}^{\rm FM}=(J/\hbar)\hat{\bf M}\cdot {\bf S}$, where $\mathbf{L}$ is the orbital angular momentum of $d$ character states in the FM, $\mathbf{S}$ is the spin, and $\hat{\bf M}$ denotes the magnetization direction of the FM. Below, we focus on the case $\hat{\mathbf{M}}=\hat{\mathbf{z}}$. At the interface, the nearest neighbor hoppings exist between the $sp$ orbitals in the NM and the $d$ orbitals in the FM. 
%he full tight-binding description of the bilayer is then obtained by combining the $sp$ and $d$ models with nearest neighbor hoppings at the NM/FM interface. 
Details of the tight-binding description are given in Ref.~\cite{Supplementary}.
All parameters of the NM and FM are set to have typical energy scales of nonmagnetic and magnetic metals. In particular, we set $\alpha_{\rm so}^{\rm FM}=100 \ \mathrm{meV}$, which is a typical SOC strength of $3d$ transition metals \cite{Wang1974, Sunko2017, Jo2018}.
We emphasize that the nonzero $\alpha_{\rm so}^{\rm FM}$ is crucial for the OT since $\hat{\mathbf{M}}$ couples only to $\mathbf{S}$ and there is no direct coupling between $\hat{\mathbf{M}}$ and $\mathbf{L}$ in the Hamiltonian. Thus for the injected orbital current to generate the OT, it should be first converted to spin through $H_\mathrm{so}^\mathrm{FM}$ and then the resulting spin can generate the torque through $H_{\rm xc}^{\rm FM}$ (Fig.~\ref{fig:schematics}).

%========================================================
%========================================================

Figure~\ref{fig:model}(b) shows the band structure of the NM/FM bilayer for $N_\mathrm{NM}=8$ and $N_\mathrm{FM}=2$, where the color represents the equilibrium expectation value of the spin-orbit correlation in the FM region $\langle \mathbf{L}\cdot\mathbf{S} \rangle_{\rm eq}^{\rm FM}$ for each state.
The correlation is negative in the lower energy range ($-1.1\ {\rm eV}<E_{n{\bf k}}<-0.7\ \mathrm{eV}$), and positive in the higher energy range ($-0.3\ {\rm eV}<E_{n{\bf k}}<+0.2\ \mathrm{eV}$). In the middle energy range ($-0.7\ {\rm eV}<E_{n{\bf k}}<-0.3\ \mathrm{eV}$), states with positive and negative correlations coexist.
We later demonstrate that $\langle \mathbf{L}\cdot\mathbf{S} \rangle_{\rm eq}^{\rm FM}$ is important for the sign of the OT.

For $\boldsymbol{\mathcal{E}}=\mathcal{E}_x\hat{\mathbf{x}}$,
we calculate the electrically generated orbital ($\mathbf{L}$) and spin ($\mathbf{S}$) accumulations at each atomic layer as a function of $z$.
From the Kubo formula, one obtains the expectation value $\left\langle \mathbf{X} (z) \right\rangle = \left\langle \mathbf{X} (z) \right\rangle^\mathrm{intra} + \left\langle \mathbf{X} (z) \right\rangle^\mathrm{inter}$ ($\mathbf{X} = \mathbf{L}$ or $\mathbf{S}$) generated by $\mathcal{E}_x$, where
\begin{subequations}
\label{eq:Kubo}
\begin{eqnarray}
\label{eq:Kubo_intra}
& &
\left\langle  \mathbf{X} (z) \right\rangle^\mathrm{intra}
=
\frac{e\hbar \mathcal{E}_x}{2\Gamma}
\sum_n \int \frac{d^2k}{(2\pi)^2} f'_{n\mathbf{k}}
\ \ \ \ \ \ \ \ \ \ \ \ \ \ \ \ \ \
\nonumber
\\
& & \ \ \ \ \
\times
\textup{Re}
\left[
\bra{u_{n\mathbf{k}}}  \mathbf{X} (z) \ket{u_{n\mathbf{k}}}
\bra{u_{n\mathbf{k}}} v_x \ket{u_{n\mathbf{k}}} \right],
\end{eqnarray}
\begin{eqnarray}
\label{eq:Kubo_inter}
& &
\left\langle   \mathbf{X} (z)  \right\rangle^\mathrm{inter}
=
-e\hbar \mathcal{E}_x
\sum_{nm} \int \frac{d^2k}{(2\pi)^2} (f_{n\mathbf{k}} - f_{m\mathbf{k}})
\nonumber
\\
& & \ \ \ \ \ \
\times
\textup{Im}
\left[
\frac{
	\bra{u_{n\mathbf{k}}} \mathbf{X} (z) \ket{u_{m\mathbf{k}}}
	\bra{u_{m\mathbf{k}}} v_x \ket{u_{n\mathbf{k}}}}
{(E_{n\mathbf{k}} - E_{m\mathbf{k}} + i\Gamma)^2}
\right],
\end{eqnarray}
\end{subequations}
are the intraband and interband contributions, respectively.
Here, $\mathbf{X}(z) = P(z) \mathbf{X} P (z)$ measures the local accumulation of $\mathbf{X}$ at $z$, where $P(z)$ is the projection operator to the atomic layer at $z$. In Eq.~(\ref{eq:Kubo}), $e>0$ is the unit charge, $\hbar$ is the Plank constant, $v_x$ is the velocity operator along the $\hat{\mathbf{x}}$ direction, $f_{n\mathbf{k}}$ is the Fermi-Dirac distribution function for a periodic part of the Bloch state $\ket{u_{n\mathbf{k}}}$ with its energy eigenvalue $E_{n\mathbf{k}}$. To incorporate the effect of disorder scatterings, we phenomenologically introduce a spectral broadening $\Gamma=25\ \mathrm{meV}$, which is a room temperature scale.

\begin{figure}[t!]
	\includegraphics[angle=0, width=0.45\textwidth]{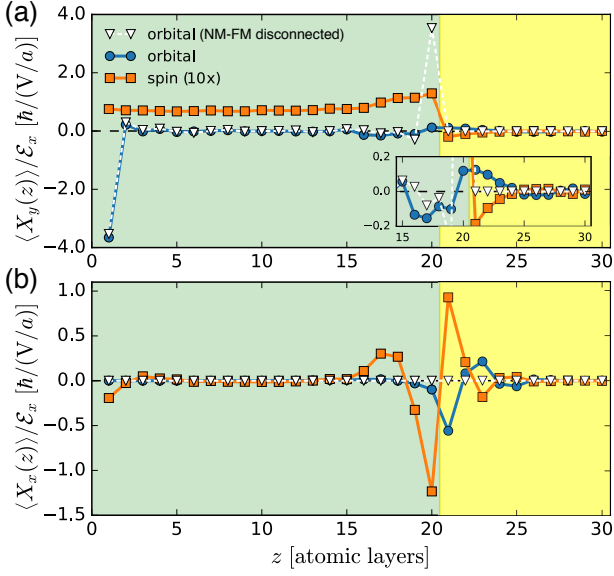}
	\caption{
(a) $\langle X_y(z) \rangle / \mathcal{E}_x$  and (b) $\langle X_x(z) \rangle / \mathcal{E}_x$ as a function of $z$ for $E_{\rm F}=-0.9\ \mathrm{eV}$. Blue circles (orange squares) depict the orbital (spin multiplied by factor 10) accumulation profile in the NM ($1\le z \le 20$) and the FM ($21 \le z \le 30$). White inverted triangles depict the orbital accumulation profile when the NM and FM are disconnected. The inset in (a) presents a magnified plot near the NM/FM interface.}
\label{fig:layer_resolved}
\end{figure}

Figure~\ref{fig:layer_resolved} shows the (a) $y$ and (b) $x$ components of the resulting $\langle \mathbf{L}(z) \rangle$ and $\langle \mathbf{S}(z) \rangle$ for the Fermi energy $E_{\rm F}=-0.9\ \mathrm{eV}$. We first consider a situation when the NM and the FM are disconnected (hoppings between the NM and FM turned off).
In the NM ($1\le z \le 20$), $\langle L_y(z)\rangle$ [white inverted triangles in Fig.~\ref{fig:layer_resolved}(a)]
has nonzero values of opposite signs at the opposite edges of the NM ($z=1$ and 20). This result can be interpreted as the orbital accumulation at the edges due to the OHE in the NM. The OHC in the NM is $\sigma_\mathrm{OH}\approx 2,000\ (\hbar/2|e|)(\Omega\cdot\mathrm{cm})^{-1}$ for $E_{\rm F}=-0.9\ \mathrm{eV}$ \cite{Go2018, Supplementary}. On the other hand, $\langle L_x(z)\rangle$ [white inverted triangles in Fig.~\ref{fig:layer_resolved}(b)] is absent in the NM. 
In the FM ($21\le z \le 30$), both $\langle L_x(z)\rangle$ and $\langle L_y(z)\rangle$ are zero, confirming the absence of the OHE in the FM. 
The spin accumulation is zero both in the NM and the FM (not shown), which is natural since the SHE is absent in both NM and FM.

Next we connect the NM and the FM (hoppings between the NM and FM turned on). Near $z=1$, which is far from the NM/FM interface, $\langle L_y(z) \rangle$ [blue circles in Fig~\ref{fig:layer_resolved}(a)] remains essentially unchanged. Near the interface ($z=20$), on the other hand, $\langle L_y(z) \rangle$ is reduced significantly  since the orbital Hall current is now injected into the FM instead of getting accumulated at the interface. 
The injected orbital Hall current in the FM produces not only $\langle L_y(z) \rangle$ but also $\langle S_y(z) \rangle$ [orange squares in Fig~\ref{fig:layer_resolved}(a) for 10$\times$ enlarged values] due to $H_{\rm so}^{\rm FM}$. 
Moreover once $\langle S_y(z)\rangle$ becomes nonzero in the FM, the spin precesses around $\hat{\mathbf{M}}$ due to $H_{\rm xc}^{\rm FM}$ and produces $\langle S_x(z) \rangle$ as well [orange squares in Fig~\ref{fig:layer_resolved}(b) for 10$\times$ enlarged values]. 
This precession results in oscillatory profiles of $\langle S_y(z)\rangle$ and $\langle S_x(z) \rangle$ in the FM, which resemble oscillatory spin accumulation profiles~\cite{Haney2013a} in a conventional situation, where a {\it spin} current is injected into a FM to generate the ST. 
%At $E_{\rm F}=-0.9\ \mathrm{eV}$, the oscillation amplitude of $\langle S_x(z) \rangle$ in the FM is about 5 larger than the corresponding amplitude of $\langle S_y(z) \rangle$ (Fig.~\ref{fig:layer_resolved}), but this feature is not universal and the ratio of the two oscillation amplitudes varies with $E_{\rm F}$. 
The oscillatory profiles of $\langle S_x(z)\rangle$ and $\langle S_y(z)\rangle$ in the FM are accompanied by similar oscillatory profiles of $\langle L_x(z)\rangle$ and $\langle L_y(z)\rangle$. The coexistence of the spin and orbital accumulation oscillations is due to $H_{\rm so}^{\rm FM}$ and we note that the spin and orbital oscillations are 180$^\circ$ out of phase for $E_{\rm F}=-0.9\ \mathrm{eV}$ (Fig.~\ref{fig:layer_resolved}), which we attribute to negative spin-orbit correlation at this energy [Fig.~\ref{fig:model}(b)]. By the way, the spin accumulation in the NM is due to partial reflection of the orbital Hall current at the NM/FM interface.

%========================================================
%========================================================

The torque $\mathbf{T}$ acting on the FM can be obtained from the spin accumulation as follows,
\begin{eqnarray}
\label{eq:OT}
\mathbf{T} = \frac{J}{\hbar} \hat{\mathbf{M}} \times
\left\langle \mathbf{S} \right\rangle^\mathrm{FM},
\end{eqnarray}
where $\left\langle \mathbf{S} \right\rangle^\mathrm{FM} = \sum_{z\in \mathrm{FM}} \left\langle \mathbf{S} (z) \right\rangle$. When the SOC of the NM is zero, $\mathbf{T}$ arises from the orbital injection and thus the resulting $\mathbf{T}$ amounts to the OT. 
Analogous to the ST, the OT can be decomposed as $\mathbf{T}=\tau_\mathrm{f} \hat{\mathbf{M}}\times\hat{\mathbf{y}}+\tau_\mathrm{d} \hat{\mathbf{M}}\times(\hat{\mathbf{M}}\times\hat{\mathbf{y}})$, where $\tau_{\mathrm{f}(\mathrm{d})}$ refers to the field(damping)-like component. When $\hat{\mathbf{M}}=\hat{\mathbf{z}}$, $\tau_\mathrm{f}=(J/\hbar) \left\langle S_y \right\rangle^\mathrm{FM}$ and $\tau_\mathrm{d}=-(J/\hbar)\left\langle S_x \right\rangle^\mathrm{FM}$.
We find that $\langle S_{y(x)}(z)\rangle$, which arises from the intraband(interband) contribution in Eq.~(\ref{eq:Kubo}), is even(odd) in $\hat{\mathbf{M}}$, thus the field(damping)-like OT is odd(even) under sign reversal of $\hat{\mathbf{M}}$. 
This is similar to the generation of the field-like and damping-like STs when a spin current polarized along $\hat{\bf y}$ direction is injected into a FM magnetized along the $\hat{\bf z}$ direction~\cite{Haney2013a}.

%========================================================
\begin{figure}[t!]
	\includegraphics[angle=0, width=0.45\textwidth]{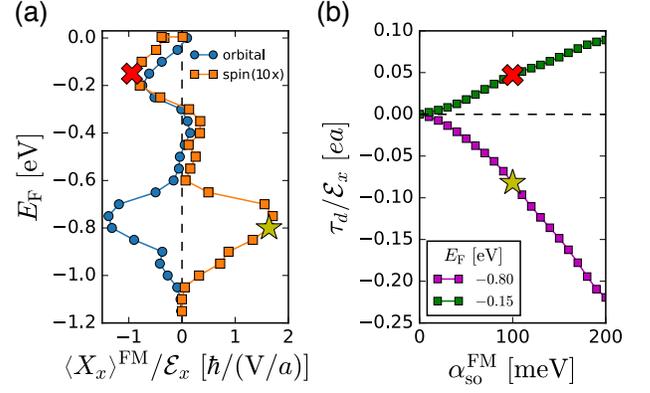}
	\caption{ \label{fig:overall_result}
(a) $\langle S_x \rangle^{\rm FM}/\mathcal{E}_x$ (orange squares for 10$\times$ magnified values) as a function of $E_{\rm F}$ with $\alpha_{\rm so}^{\rm FM}=100$ meV. For comparison, $\langle L_x \rangle^{\rm FM}/\mathcal{E}_x$ (blue circles) is also shown.
%Damping-like OT $\tau_{\rm d}$ (a) as a function of $E_{\rm F}$ with the SOC parameter $\alpha_{\rm so}^{\rm FM}$ in the FM fixed to 50 meV and 
(b) $\tau_{\rm d}/\mathcal{E}_x$ as a function of $\alpha_{\rm so}^{\rm FM}$ with $E_{\rm F}=-0.80\ \mathrm{eV}$ (purple squares) and $E_{\rm F}=-0.15\ \mathrm{eV}$ (green squares). The yellow star and red cross symbols in (a) and (b) are obtained for the same $E_{\rm F}$ and $\alpha_{\rm so}^{\rm FM}$. For this calculation, smaller system size is used ($N_{\rm NM}=8$ and $N_{\rm FM}=2$).
%		Layer resolved plots for the electric responses of (a) $X_y$ and (b) $X_x$ at $E_\mathrm{F}=-0.9\ \mathrm{eV}$, where $\mathbf{X}=\mathbf{L}$ (blue  circle) and $\mathbf{S}$ (orange square). The green and yellow regions indicate the NM and FM, respectively.
}
\end{figure}
%========================================================

Since $\tau_\mathrm{d}$ plays a more important role for the current-induced magnetization dynamics than $\tau_\mathrm{f}$~\cite{Liu2012a,Liu2012b}, we focus on $\langle S_x\rangle^{\rm FM}$. A result for $\langle S_y\rangle^{\rm FM}$ is given in Ref.~\cite{Supplementary}.
Figure~\ref{fig:overall_result}(a) shows that the ratio
$\langle S_x\rangle^{\rm FM}/\mathcal{E}_x$ (orange squares) is positive for $-1.0\ \mathrm{eV} \lesssim E_{\rm F} \lesssim -0.6\ \mathrm{eV}$ and negative for $-0.2\ \mathrm{eV} \lesssim E_{\rm F} \lesssim 0.0\ \mathrm{eV}$. 
For comparison, the ratio $\langle L_x\rangle^{\rm FM}/\mathcal{E}_x$ (blue circles) is also shown, where $\langle L_x\rangle^{\rm FM}\equiv\sum_{z\in {\rm FM}}\langle L_x(z)\rangle$.
Note that the relative ratio between $\langle S_x\rangle^{\rm FM}$ and $\langle L_x\rangle^{\rm FM}$ is negative for $-1.0\ \mathrm{eV} \lesssim E_{\rm F} \lesssim -0.6\ \mathrm{eV}$, and positive for $-0.2\ \mathrm{eV} \lesssim E_{\rm F} \lesssim 0.0\ \mathrm{eV}$. 
The $E_{\rm F}$-dependence of the relative ratio sign closely resembles the energy dependence of the spin-orbit correlation $\langle {\bf L}\cdot {\bf S}\rangle_{\rm eq}^{\rm FM}$ in Fig.~\ref{fig:model}(b). By combining the calculation result in Fig.~\ref{fig:overall_result}(a) with the fact that the OHC of the NM is positive essentially for all $E_{\rm F}$ \cite{Go2018}, we find that the sign of $\langle S_x\rangle^{\rm FM}/\mathcal{E}_x$ tends to be determined by the sign of the product between the OHC of the NM and the spin-orbit correlation in the FM. Considering that $\langle S_x\rangle^{\rm FM}$ determines the damping-like OT, the latter tendency may be regarded as the OT counterpart of the sign ``rule" for the ST; the damping-like ST tends to be determined by the sign of the spin Hall conductivity (SHC) in the NM~\cite{Liu2012a,Liu2012b, Haney2013a}.

Figure~\ref{fig:overall_result}(b) shows the ratio $\tau_{\rm d}/\mathcal{E}_x$ as a function of $\alpha_{\rm so}^{\rm FM}$ for $E_{\rm F}=-0.80\ \mathrm{eV}$ (purple squares) and $E_\mathrm{F}=-0.15\ \mathrm{eV}$ (green squares). These two $E_{\rm F}$ values are close to the peak positions in Fig.~\ref{fig:overall_result}(a) (denoted by the yellow star and the red cross).
For these favorable choices of $E_{\rm F}$, values of $\tau_{\rm d}/\mathcal{E}_x$ are $-0.08\ ea$ and $+0.05\ ea$ for $\alpha_{\rm so}^{\rm FM}=100\ \mathrm{meV}$, which is SOC energy scale for $3d$ FMs. Here $a$ is the lattice constant. By increasing $\alpha_{\rm so}^{\rm FM}$, they reach up to $-0.22\ ea$ and $+0.09\ ea$ for $\alpha_{\rm so}^{\rm FM}=200\ \mathrm{meV}$, which is SOC energy scale for $4d$ transition metals. 
Note that these values $\tau_{\rm d}/\mathcal{E}_x \sim 0.1\ ea$ for $\alpha_\mathrm{so}^\mathrm{FM}=100,\ 200\ \mathrm{meV}$ are not negligible compared to the corresponding value $\sim 0.5\ ea$ for the damping-like torque calculated for the Pt/Co bilayer~\cite{Freimuth2010,Mahfouzi2018} with the SOC strength of $500\ \mathrm{meV}$ for Pt. Then, considering that the OHC in real materials such as V is gigantic $\sigma_\mathrm{OH} \sim 12,000\ (\hbar/2|e|)(\Omega\cdot\mathrm{cm})^{-1}$, which is about 6 times larger than the OHC of the $sp$ model used in our calculation, $\tau_{\rm d}/\mathcal{E}_x$ for real NMs may be proportionally larger and comparable to the corresponding ST value for the Pt/Co bilayer. Although quantitative predictions on $\tau_{\rm d}/\mathcal{E}_x$ require realistic calculations that take material details into account, we argue it is still reasonable to expect that the OT may be sizable for a FM with weak SOC, thus providing an alternative route to enhancing the torque efficiency.

%%%%%%%%%%%%%%%%%
\begin{table}[t!]
\centering
\begin{tabular}{|c|c|c|} \hline
& $\langle {\bf L}\cdot{\bf S}\rangle_{\rm eq}^{\rm FM}>0$ & $\langle {\bf L}\cdot{\bf S}\rangle_{\rm eq}^{\rm FM}<0$ \\ \hline
$\langle {\bf L}\cdot{\bf S}\rangle_{\rm eq}^{\rm NM}>0$ & same sign & opposite signs \\ \hline
$\langle {\bf L}\cdot{\bf S}\rangle_{\rm eq}^{\rm NM}<0$ & opposite signs & same sign \\ \hline
\end{tabular}
\caption{Relative signs of the OT and ST depending on the spin-orbit correlations in the NM and FM.}
\label{tab:OT_ST}
\end{table}
%%%%%%%%%%%%%%%%%

% ********************* DISCUSSION *********************

So far we have assumed the SOC is absent in the NM. Now we consider a situation where not only the FM but also the NM have the SOC. Thus the $sp$ model Hamiltonian for the NM now includes
\begin{equation}
H_{\rm so}^{\rm NM}=\frac{\alpha_{\rm so}^{\rm NM}}{\hbar^2}{\bf L}\cdot{\bf S},
\label{eq:SOC_NM}
\end{equation}
where $\alpha_\mathrm{so}^\mathrm{NM}$ is the SOC parameter in the NM. Since $s$ character states do not carry the orbital angular momentum, $\mathbf{L}$ in Eq.~\eqref{eq:SOC_NM} acts only on $p$ character states.
Due to $H_{\rm so}^{\rm NM}$, the NM exhibits SHE, as well as the OHE. Thus, on top of the OT, injection of the spin Hall current into the FM generates the ST. 
It is known that OHE and SHE occur in the same(opposite) direction if $\langle {\bf L}\cdot {\bf S} \rangle_{\rm eq}^{\rm NM}$ is positive(negative) at $E_\mathrm{F}$ \cite{Tanaka2008, Kontani2009, Go2018}. Thus, when $\langle {\bf L}\cdot {\bf S} \rangle_{\rm eq}^{\rm FM}>0$ at $E_\mathrm{F}$, which is a case for Ni, the OT and ST add up if $\langle {\bf L}\cdot {\bf S} \rangle_{\rm eq}^{\rm NM}>0$ and cancel each other if $\langle {\bf L}\cdot {\bf S} \rangle_{\rm eq}^{\rm NM}<0$. This situation becomes the opposite when $\langle {\bf L}\cdot {\bf S} \rangle_{\rm eq}^{\rm FM}<0$, as in Gd. The result is summarized in Table~\ref{tab:OT_ST}, which is supported by our numerical calculation~\cite{Supplementary}. This implies that the total torque may go even beyond the level expected from the theoretical value for the SHC of a NM, as in recent experiments \cite{Zhu2018, Du2018}
%
%Recent torque measurement experiments on Au$_{0.25}$Pt$_{0.75}$/Co~\cite{Zhu2018} and epitaxial Pt/Co \cite{Du2018} may be this case.
When the OT and ST cancel each other, the total torque may even exhibit the opposite sign compared to the sign expected from the SHC of the NM. For example, Ta and W exhibit the opposite signs of the OHC and SHC \cite{Tanaka2008}.

Unfortunately, the OT and ST exhibit \emph{qualitatively} similar behavior, thus disentangling the OT from the ST is challenging. The orbital and spin operator transform in the same way for symmetry operations, i.e. both OT and ST exhibit the same angular dependence. Nonetheless, the OT and ST are expected to exhibit different \emph{quantitative} features. One characteristic feature of the OT is its strong correlation with $\left\langle \mathbf{L}\cdot\mathbf{S}\right\rangle_\mathrm{eq}^\mathrm{FM}$ as demonstrated in Figs. \ref{fig:layer_resolved} and \ref{fig:overall_result}(b). This suggests that the OT can be probed through material variation of a FM. This is in stark contrast to the ST, where the a role of the FM is less important. We expect that alloying a FM with heavy elements would not only increase the OT since the conversion from the orbital to spin becomes more efficient but also provide a way to systematically tune the spin-orbit correlation of the FM \cite{FM_SHE_comment}.

Another distinct feature of the OT compared to the ST is its dependence on the interface crystallinity. For the orbital injection across the interface, it must occur through orbital hybridizations at the NM/FM interface. In the tight-binding model in Fig. ~\ref{fig:model}(a), $p_z^\mathrm{NM}-d_{z^2-x^2}^\mathrm{FM}$ and $p_x^\mathrm{NM}-d_{zx}^\mathrm{FM}$ hybridizations are crucial for transferring $\left\langle L_y \right\rangle$ \cite{Supplementary}. Thus even spin-conserving interface scatterings can result in orbital relaxation, making the \emph{orbital transparency} more sensitive to the interface crystallinity than the spin transparency. Particularly, when the NM and FM elements tend to be mixed, the OT will be suppressed since the atomic ordering of the NM and FM atoms disappears at the interface \cite{atomic_ordering_comment}.

%This tendency may provide experimental method to distinguish the OT from the ST. For instance, in bilayers with the same sign of the OT and the ST, the NM/FM interface quality enhancement will increase the OT and the total torque. Sizable increase of the total torque has been reported for epitaxial Pt/Co~\cite{Du2018}. On the other hand, in bilayers with opposite signs of the OT and the ST, the interface quality enhancement will increase the OT and {\it reduce} the total torque. 

When the NM/FM bilayer consists only of light elements with weak SOC, the total torque is dominated by the OT as in Fig.~\ref{fig:schematics}, which is advantageous for unambiguously quantifying the OT. In the past, unexpectedly large torque was measured in samples containing Cr \cite{Du2014, Qu2015} and Py \cite{Miao2013, Tsukahara2014} in spite of small SOC of these elements. These results may be related to the OT, which requires further investigation. Finally, we remark that the orbital angular momentum can be generated not only by the OHE but also by the interfacial Rashba-type states \cite{Go2017,Chen2018}, which may be related to sizable field-like torque measured in Py/(Cu)/$\mathrm{AlO}_x$ structure~\cite{Emori2016}.

%surface-oxidized Cu/Py~\cite{An2016, Gao2018}, and (Py,Cu)/$\mathrm{AlO}_x$~\cite{Emori2016}. These experimental results may be related to the OT.

%The OT may be probed also through material variation of a FM. This is in clear contrast to present experimental efforts focused on the material variation of a NM while the material choice for a FM is mostly fixed to $3d$ FMs with weak SOC. According to Fig.~\ref{fig:overall_result}(b), the OT is enhanced in FMs with strong SOC (Gd for instance). Then in bilayers with opposite signs of the OT and the ST, the total torque may even have a sign that is opposite to the sign expected from the SHC sign of a NM.

%In summary, we have demonstrated torque generation by the orbital injection. Compared to conventional ST utilizing the SHE of NMs with strong SOC, the OT resulting from the OHE of NMs with weak SOC may be comparable in magnitude since the OHCs can be larger than the SHCs. The OT may be responsible for experimental results that are difficult to explain by the SHE. Some additional experiments to probe the OT are suggested. The OT not only provides a way to probe the OHE experimentally but also widens material choices for developing spintronic devices.

\
\

\let\oldaddcontentsline\addcontentsline% Store \addcontentsline
\renewcommand{\addcontentsline}[3]{}% Make \addcontentsline a no-op

% Acknowledgement
\begin{acknowledgments}
We acknowledge Daegeun Jo, Junyeon Kim, YoshiChika Otani, Jan-Philipp Hanke, Frank Freimuth, Yuriy Mokrousov, and Kyung-Jin Lee for insightful discussions. D. G. and H.-W. L. were supported by the
%Samsung Science \& Technology Foundation
SSTF (Grant No. BA-1501-07).
\end{acknowledgments}

\bibliography{bib_orbital_torque}

\begin{thebibliography}{40}%
\makeatletter
\providecommand \@ifxundefined [1]{%
 \@ifx{#1\undefined}
}%
\providecommand \@ifnum [1]{%
 \ifnum #1\expandafter \@firstoftwo
 \else \expandafter \@secondoftwo
 \fi
}%
\providecommand \@ifx [1]{%
 \ifx #1\expandafter \@firstoftwo
 \else \expandafter \@secondoftwo
 \fi
}%
\providecommand \natexlab [1]{#1}%
\providecommand \enquote  [1]{``#1''}%
\providecommand \bibnamefont  [1]{#1}%
\providecommand \bibfnamefont [1]{#1}%
\providecommand \citenamefont [1]{#1}%
\providecommand \href@noop [0]{\@secondoftwo}%
\providecommand \href [0]{\begingroup \@sanitize@url \@href}%
\providecommand \@href[1]{\@@startlink{#1}\@@href}%
\providecommand \@@href[1]{\endgroup#1\@@endlink}%
\providecommand \@sanitize@url [0]{\catcode `\\12\catcode `\$12\catcode
  `\&12\catcode `\#12\catcode `\^12\catcode `\_12\catcode `\%12\relax}%
\providecommand \@@startlink[1]{}%
\providecommand \@@endlink[0]{}%
\providecommand \url  [0]{\begingroup\@sanitize@url \@url }%
\providecommand \@url [1]{\endgroup\@href {#1}{\urlprefix }}%
\providecommand \urlprefix  [0]{URL }%
\providecommand \Eprint [0]{\href }%
\providecommand \doibase [0]{http://dx.doi.org/}%
\providecommand \selectlanguage [0]{\@gobble}%
\providecommand \bibinfo  [0]{\@secondoftwo}%
\providecommand \bibfield  [0]{\@secondoftwo}%
\providecommand \translation [1]{[#1]}%
\providecommand \BibitemOpen [0]{}%
\providecommand \bibitemStop [0]{}%
\providecommand \bibitemNoStop [0]{.\EOS\space}%
\providecommand \EOS [0]{\spacefactor3000\relax}%
\providecommand \BibitemShut  [1]{\csname bibitem#1\endcsname}%
\let\auto@bib@innerbib\@empty
%</preamble>
\bibitem [{\citenamefont {Pi}\ \emph {et~al.}(2010)\citenamefont {Pi},
  \citenamefont {Won~Kim}, \citenamefont {Bae}, \citenamefont {Lee},
  \citenamefont {Cho}, \citenamefont {Kim},\ and\ \citenamefont
  {Seo}}]{Pi2010}%
  \BibitemOpen
  \bibfield  {author} {\bibinfo {author} {\bibfnamefont {Ung~Hwan}\
  \bibnamefont {Pi}}, \bibinfo {author} {\bibfnamefont {Kee}\ \bibnamefont
  {Won~Kim}}, \bibinfo {author} {\bibfnamefont {Ji~Young}\ \bibnamefont {Bae}},
  \bibinfo {author} {\bibfnamefont {Sung~Chul}\ \bibnamefont {Lee}}, \bibinfo
  {author} {\bibfnamefont {Young~Jin}\ \bibnamefont {Cho}}, \bibinfo {author}
  {\bibfnamefont {Kwang~Seok}\ \bibnamefont {Kim}}, \ and\ \bibinfo {author}
  {\bibfnamefont {Sunae}\ \bibnamefont {Seo}},\ }\bibfield  {title} {\enquote
  {\bibinfo {title} {{Tilting of the spin orientation induced by Rashba effect
  in ferromagnetic metal layer}},}\ }\href@noop {} {\bibfield  {journal}
  {\bibinfo  {journal} {Appl. Phys. Lett.}\ }\textbf {\bibinfo {volume} {97}},\
  \bibinfo {pages} {162507} (\bibinfo {year} {2010})}\BibitemShut {NoStop}%
\bibitem [{\citenamefont {Mihai~Miron}\ \emph {et~al.}(2010)\citenamefont
  {Mihai~Miron}, \citenamefont {Gaudin}, \citenamefont {Auffret}, \citenamefont
  {Rodmacq}, \citenamefont {Schuhl}, \citenamefont {Pizzini}, \citenamefont
  {Vogel},\ and\ \citenamefont {Gambardella}}]{Miron2010}%
  \BibitemOpen
  \bibfield  {author} {\bibinfo {author} {\bibfnamefont {Ioan}\ \bibnamefont
  {Mihai~Miron}}, \bibinfo {author} {\bibfnamefont {Gilles}\ \bibnamefont
  {Gaudin}}, \bibinfo {author} {\bibfnamefont {St{\'e}phane}\ \bibnamefont
  {Auffret}}, \bibinfo {author} {\bibfnamefont {Bernard}\ \bibnamefont
  {Rodmacq}}, \bibinfo {author} {\bibfnamefont {Alain}\ \bibnamefont {Schuhl}},
  \bibinfo {author} {\bibfnamefont {Stefania}\ \bibnamefont {Pizzini}},
  \bibinfo {author} {\bibfnamefont {Jan}\ \bibnamefont {Vogel}}, \ and\
  \bibinfo {author} {\bibfnamefont {Pietro}\ \bibnamefont {Gambardella}},\
  }\bibfield  {title} {\enquote {\bibinfo {title} {Current-driven spin torque
  induced by the rashba effect in a ferromagnetic metal layer},}\ }\href@noop
  {} {\bibfield  {journal} {\bibinfo  {journal} {Nat. Mater.}\ }\textbf
  {\bibinfo {volume} {9}},\ \bibinfo {pages} {230} (\bibinfo {year}
  {2010})}\BibitemShut {NoStop}%
\bibitem [{\citenamefont {Miron}\ \emph
  {et~al.}(2011{\natexlab{a}})\citenamefont {Miron}, \citenamefont {Moore},
  \citenamefont {Szambolics}, \citenamefont {Buda-Prejbeanu}, \citenamefont
  {Auffret}, \citenamefont {Rodmacq}, \citenamefont {Pizzini}, \citenamefont
  {Vogel}, \citenamefont {Bonfim}, \citenamefont {Schuhl},\ and\ \citenamefont
  {Gaudin}}]{Miron2011a}%
  \BibitemOpen
  \bibfield  {author} {\bibinfo {author} {\bibfnamefont {Ioan~Mihai}\
  \bibnamefont {Miron}}, \bibinfo {author} {\bibfnamefont {Thomas}\
  \bibnamefont {Moore}}, \bibinfo {author} {\bibfnamefont {Helga}\ \bibnamefont
  {Szambolics}}, \bibinfo {author} {\bibfnamefont {Liliana~Daniela}\
  \bibnamefont {Buda-Prejbeanu}}, \bibinfo {author} {\bibfnamefont
  {St{\'e}phane}\ \bibnamefont {Auffret}}, \bibinfo {author} {\bibfnamefont
  {Bernard}\ \bibnamefont {Rodmacq}}, \bibinfo {author} {\bibfnamefont
  {Stefania}\ \bibnamefont {Pizzini}}, \bibinfo {author} {\bibfnamefont {Jan}\
  \bibnamefont {Vogel}}, \bibinfo {author} {\bibfnamefont {Marlio}\
  \bibnamefont {Bonfim}}, \bibinfo {author} {\bibfnamefont {Alain}\
  \bibnamefont {Schuhl}}, \ and\ \bibinfo {author} {\bibfnamefont {Gilles}\
  \bibnamefont {Gaudin}},\ }\bibfield  {title} {\enquote {\bibinfo {title}
  {Fast current-induced domain-wall motion controlled by the rashba effect},}\
  }\href@noop {} {\bibfield  {journal} {\bibinfo  {journal} {Mat. Mater.}\
  }\textbf {\bibinfo {volume} {10}},\ \bibinfo {pages} {419} (\bibinfo {year}
  {2011}{\natexlab{a}})}\BibitemShut {NoStop}%
\bibitem [{\citenamefont {Miron}\ \emph
  {et~al.}(2011{\natexlab{b}})\citenamefont {Miron}, \citenamefont {Garello},
  \citenamefont {Gaudin}, \citenamefont {Zermatten}, \citenamefont {Costache},
  \citenamefont {Auffret}, \citenamefont {Bandiera}, \citenamefont {Rodmacq},
  \citenamefont {Schuhl},\ and\ \citenamefont {Gambardella}}]{Miron2011b}%
  \BibitemOpen
  \bibfield  {author} {\bibinfo {author} {\bibfnamefont {Ioan~Mihai}\
  \bibnamefont {Miron}}, \bibinfo {author} {\bibfnamefont {Kevin}\ \bibnamefont
  {Garello}}, \bibinfo {author} {\bibfnamefont {Gilles}\ \bibnamefont
  {Gaudin}}, \bibinfo {author} {\bibfnamefont {Pierre-Jean}\ \bibnamefont
  {Zermatten}}, \bibinfo {author} {\bibfnamefont {Marius~V.}\ \bibnamefont
  {Costache}}, \bibinfo {author} {\bibfnamefont {St{\'e}phane}\ \bibnamefont
  {Auffret}}, \bibinfo {author} {\bibfnamefont {S{\'e}bastien}\ \bibnamefont
  {Bandiera}}, \bibinfo {author} {\bibfnamefont {Bernard}\ \bibnamefont
  {Rodmacq}}, \bibinfo {author} {\bibfnamefont {Alain}\ \bibnamefont {Schuhl}},
  \ and\ \bibinfo {author} {\bibfnamefont {Pietro}\ \bibnamefont
  {Gambardella}},\ }\bibfield  {title} {\enquote {\bibinfo {title}
  {Perpendicular switching of a single ferromagnetic layer induced by in-plane
  current injection},}\ }\href {http://dx.doi.org/10.1038/nature10309}
  {\bibfield  {journal} {\bibinfo  {journal} {Nature (London)}\ }\textbf
  {\bibinfo {volume} {476}},\ \bibinfo {pages} {189} (\bibinfo {year}
  {2011}{\natexlab{b}})}\BibitemShut {NoStop}%
\bibitem [{\citenamefont {Pai}\ \emph {et~al.}(2012)\citenamefont {Pai},
  \citenamefont {Liu}, \citenamefont {Li}, \citenamefont {Tseng}, \citenamefont
  {Ralph},\ and\ \citenamefont {Buhrman}}]{Pai2012}%
  \BibitemOpen
  \bibfield  {author} {\bibinfo {author} {\bibfnamefont {Chi-Feng}\
  \bibnamefont {Pai}}, \bibinfo {author} {\bibfnamefont {Luqiao}\ \bibnamefont
  {Liu}}, \bibinfo {author} {\bibfnamefont {Y.}~\bibnamefont {Li}}, \bibinfo
  {author} {\bibfnamefont {H.~W.}\ \bibnamefont {Tseng}}, \bibinfo {author}
  {\bibfnamefont {D.~C.}\ \bibnamefont {Ralph}}, \ and\ \bibinfo {author}
  {\bibfnamefont {R.~A.}\ \bibnamefont {Buhrman}},\ }\bibfield  {title}
  {\enquote {\bibinfo {title} {{Spin transfer torque devices utilizing the
  giant spin Hall effect of tungsten}},}\ }\href@noop {} {\bibfield  {journal}
  {\bibinfo  {journal} {Applied Physics Letters}\ }\textbf {\bibinfo {volume}
  {101}},\ \bibinfo {pages} {122404} (\bibinfo {year} {2012})}\BibitemShut
  {NoStop}%
\bibitem [{\citenamefont {Liu}\ \emph {et~al.}(2012{\natexlab{a}})\citenamefont
  {Liu}, \citenamefont {Lee}, \citenamefont {Gudmundsen}, \citenamefont
  {Ralph},\ and\ \citenamefont {Buhrman}}]{Liu2012a}%
  \BibitemOpen
  \bibfield  {author} {\bibinfo {author} {\bibfnamefont {Luqiao}\ \bibnamefont
  {Liu}}, \bibinfo {author} {\bibfnamefont {O.~J.}\ \bibnamefont {Lee}},
  \bibinfo {author} {\bibfnamefont {T.~J.}\ \bibnamefont {Gudmundsen}},
  \bibinfo {author} {\bibfnamefont {D.~C.}\ \bibnamefont {Ralph}}, \ and\
  \bibinfo {author} {\bibfnamefont {R.~A.}\ \bibnamefont {Buhrman}},\
  }\bibfield  {title} {\enquote {\bibinfo {title} {{Current-Induced Switching
  of Perpendicularly Magnetized Magnetic Layers Using Spin Torque from the Spin
  Hall Effect}},}\ }\href {\doibase 10.1103/PhysRevLett.109.096602} {\bibfield
  {journal} {\bibinfo  {journal} {Phys. Rev. Lett.}\ }\textbf {\bibinfo
  {volume} {109}},\ \bibinfo {pages} {096602} (\bibinfo {year}
  {2012}{\natexlab{a}})}\BibitemShut {NoStop}%
\bibitem [{\citenamefont {Liu}\ \emph {et~al.}(2012{\natexlab{b}})\citenamefont
  {Liu}, \citenamefont {Pai}, \citenamefont {Li}, \citenamefont {Tseng},
  \citenamefont {Ralph},\ and\ \citenamefont {Buhrman}}]{Liu2012b}%
  \BibitemOpen
  \bibfield  {author} {\bibinfo {author} {\bibfnamefont {Luqiao}\ \bibnamefont
  {Liu}}, \bibinfo {author} {\bibfnamefont {Chi-Feng}\ \bibnamefont {Pai}},
  \bibinfo {author} {\bibfnamefont {Y.}~\bibnamefont {Li}}, \bibinfo {author}
  {\bibfnamefont {H.~W.}\ \bibnamefont {Tseng}}, \bibinfo {author}
  {\bibfnamefont {D.~C.}\ \bibnamefont {Ralph}}, \ and\ \bibinfo {author}
  {\bibfnamefont {R.~A.}\ \bibnamefont {Buhrman}},\ }\bibfield  {title}
  {\enquote {\bibinfo {title} {{Spin-Torque Switching with the Giant Spin Hall
  Effect of Tantalum}},}\ }\href {\doibase 10.1126/science.1218197} {\bibfield
  {journal} {\bibinfo  {journal} {Science}\ }\textbf {\bibinfo {volume}
  {336}},\ \bibinfo {pages} {555--558} (\bibinfo {year}
  {2012}{\natexlab{b}})}\BibitemShut {NoStop}%
\bibitem [{\citenamefont {Kim}\ \emph {et~al.}(2012)\citenamefont {Kim},
  \citenamefont {Seo}, \citenamefont {Ryu}, \citenamefont {Lee},\ and\
  \citenamefont {Lee}}]{Kim2012}%
  \BibitemOpen
  \bibfield  {author} {\bibinfo {author} {\bibfnamefont {Kyoung-Whan}\
  \bibnamefont {Kim}}, \bibinfo {author} {\bibfnamefont {Soo-Man}\ \bibnamefont
  {Seo}}, \bibinfo {author} {\bibfnamefont {Jisu}\ \bibnamefont {Ryu}},
  \bibinfo {author} {\bibfnamefont {Kyung-Jin}\ \bibnamefont {Lee}}, \ and\
  \bibinfo {author} {\bibfnamefont {Hyun-Woo}\ \bibnamefont {Lee}},\ }\bibfield
   {title} {\enquote {\bibinfo {title} {Magnetization dynamics induced by
  in-plane currents in ultrathin magnetic nanostructures with rashba spin-orbit
  coupling},}\ }\href {\doibase 10.1103/PhysRevB.85.180404} {\bibfield
  {journal} {\bibinfo  {journal} {Phys. Rev. B}\ }\textbf {\bibinfo {volume}
  {85}},\ \bibinfo {pages} {180404(R)} (\bibinfo {year} {2012})}\BibitemShut
  {NoStop}%
\bibitem [{\citenamefont {Haney}\ \emph
  {et~al.}(2013{\natexlab{a}})\citenamefont {Haney}, \citenamefont {Lee},
  \citenamefont {Lee}, \citenamefont {Manchon},\ and\ \citenamefont
  {Stiles}}]{Haney2013a}%
  \BibitemOpen
  \bibfield  {author} {\bibinfo {author} {\bibfnamefont {Paul~M.}\ \bibnamefont
  {Haney}}, \bibinfo {author} {\bibfnamefont {Hyun-Woo}\ \bibnamefont {Lee}},
  \bibinfo {author} {\bibfnamefont {Kyung-Jin}\ \bibnamefont {Lee}}, \bibinfo
  {author} {\bibfnamefont {Aur\'elien}\ \bibnamefont {Manchon}}, \ and\
  \bibinfo {author} {\bibfnamefont {M.~D.}\ \bibnamefont {Stiles}},\ }\bibfield
   {title} {\enquote {\bibinfo {title} {Current induced torques and interfacial
  spin-orbit coupling: Semiclassical modeling},}\ }\href {\doibase
  10.1103/PhysRevB.87.174411} {\bibfield  {journal} {\bibinfo  {journal} {Phys.
  Rev. B}\ }\textbf {\bibinfo {volume} {87}},\ \bibinfo {pages} {174411}
  (\bibinfo {year} {2013}{\natexlab{a}})}\BibitemShut {NoStop}%
\bibitem [{\citenamefont {Haney}\ \emph
  {et~al.}(2013{\natexlab{b}})\citenamefont {Haney}, \citenamefont {Lee},
  \citenamefont {Lee}, \citenamefont {Manchon},\ and\ \citenamefont
  {Stiles}}]{Haney2013b}%
  \BibitemOpen
  \bibfield  {author} {\bibinfo {author} {\bibfnamefont {Paul~M.}\ \bibnamefont
  {Haney}}, \bibinfo {author} {\bibfnamefont {Hyun-Woo}\ \bibnamefont {Lee}},
  \bibinfo {author} {\bibfnamefont {Kyung-Jin}\ \bibnamefont {Lee}}, \bibinfo
  {author} {\bibfnamefont {Aur\'elien}\ \bibnamefont {Manchon}}, \ and\
  \bibinfo {author} {\bibfnamefont {M.~D.}\ \bibnamefont {Stiles}},\ }\bibfield
   {title} {\enquote {\bibinfo {title} {Current-induced torques and interfacial
  spin-orbit coupling},}\ }\href {\doibase 10.1103/PhysRevB.88.214417}
  {\bibfield  {journal} {\bibinfo  {journal} {Phys. Rev. B}\ }\textbf {\bibinfo
  {volume} {88}},\ \bibinfo {pages} {214417} (\bibinfo {year}
  {2013}{\natexlab{b}})}\BibitemShut {NoStop}%
\bibitem [{\citenamefont {Garello}\ \emph {et~al.}(2013)\citenamefont
  {Garello}, \citenamefont {Miron}, \citenamefont {Avci}, \citenamefont
  {Freimuth}, \citenamefont {Mokrousov}, \citenamefont {Bl{\"u}gel},
  \citenamefont {Auffret}, \citenamefont {Boulle}, \citenamefont {Gaudin},\
  and\ \citenamefont {Gambardella}}]{Garello2013}%
  \BibitemOpen
  \bibfield  {author} {\bibinfo {author} {\bibfnamefont {Kevin}\ \bibnamefont
  {Garello}}, \bibinfo {author} {\bibfnamefont {Ioan~Mihai}\ \bibnamefont
  {Miron}}, \bibinfo {author} {\bibfnamefont {Can~Onur}\ \bibnamefont {Avci}},
  \bibinfo {author} {\bibfnamefont {Frank}\ \bibnamefont {Freimuth}}, \bibinfo
  {author} {\bibfnamefont {Yuriy}\ \bibnamefont {Mokrousov}}, \bibinfo {author}
  {\bibfnamefont {Stefan}\ \bibnamefont {Bl{\"u}gel}}, \bibinfo {author}
  {\bibfnamefont {St{\'e}phane}\ \bibnamefont {Auffret}}, \bibinfo {author}
  {\bibfnamefont {Olivier}\ \bibnamefont {Boulle}}, \bibinfo {author}
  {\bibfnamefont {Gilles}\ \bibnamefont {Gaudin}}, \ and\ \bibinfo {author}
  {\bibfnamefont {Pietro}\ \bibnamefont {Gambardella}},\ }\bibfield  {title}
  {\enquote {\bibinfo {title} {Symmetry and magnitude of spin-orbit torques in
  ferromagnetic heterostructures},}\ }\href@noop {} {\bibfield  {journal}
  {\bibinfo  {journal} {Nat. Nanotechnol.}\ }\textbf {\bibinfo {volume} {8}},\
  \bibinfo {pages} {587} (\bibinfo {year} {2013})}\BibitemShut {NoStop}%
\bibitem [{\citenamefont {Garello}\ \emph {et~al.}(2014)\citenamefont
  {Garello}, \citenamefont {Avci}, \citenamefont {Miron}, \citenamefont
  {Baumgartner}, \citenamefont {Ghosh}, \citenamefont {Auffret}, \citenamefont
  {Boulle}, \citenamefont {Gaudin},\ and\ \citenamefont
  {Gambardella}}]{Garello2014}%
  \BibitemOpen
  \bibfield  {author} {\bibinfo {author} {\bibfnamefont {Kevin}\ \bibnamefont
  {Garello}}, \bibinfo {author} {\bibfnamefont {Can~Onur}\ \bibnamefont
  {Avci}}, \bibinfo {author} {\bibfnamefont {Ioan~Mihai}\ \bibnamefont
  {Miron}}, \bibinfo {author} {\bibfnamefont {Manuel}\ \bibnamefont
  {Baumgartner}}, \bibinfo {author} {\bibfnamefont {Abhijit}\ \bibnamefont
  {Ghosh}}, \bibinfo {author} {\bibfnamefont {Stéphane}\ \bibnamefont
  {Auffret}}, \bibinfo {author} {\bibfnamefont {Olivier}\ \bibnamefont
  {Boulle}}, \bibinfo {author} {\bibfnamefont {Gilles}\ \bibnamefont {Gaudin}},
  \ and\ \bibinfo {author} {\bibfnamefont {Pietro}\ \bibnamefont
  {Gambardella}},\ }\bibfield  {title} {\enquote {\bibinfo {title} {Ultrafast
  magnetization switching by spin-orbit torques},}\ }\href@noop {} {\bibfield
  {journal} {\bibinfo  {journal} {Applied Physics Letters}\ }\textbf {\bibinfo
  {volume} {105}},\ \bibinfo {pages} {212402} (\bibinfo {year}
  {2014})}\BibitemShut {NoStop}%
\bibitem [{\citenamefont {Kurebayashi}\ \emph {et~al.}(2014)\citenamefont
  {Kurebayashi}, \citenamefont {Sinova}, \citenamefont {Fang}, \citenamefont
  {Irvine}, \citenamefont {Skinner}, \citenamefont {Wunderlich}, \citenamefont
  {Nov{\'a}k}, \citenamefont {Campion}, \citenamefont {Gallagher},
  \citenamefont {Vehstedt}, \citenamefont {Z{\^a}rbo}, \citenamefont
  {V{\'y}born{\'y}}, \citenamefont {Ferguson},\ and\ \citenamefont
  {Jungwirth}}]{Kurebayashi2014}%
  \BibitemOpen
  \bibfield  {author} {\bibinfo {author} {\bibfnamefont {H.}~\bibnamefont
  {Kurebayashi}}, \bibinfo {author} {\bibfnamefont {Jairo}\ \bibnamefont
  {Sinova}}, \bibinfo {author} {\bibfnamefont {D.}~\bibnamefont {Fang}},
  \bibinfo {author} {\bibfnamefont {A.~C.}\ \bibnamefont {Irvine}}, \bibinfo
  {author} {\bibfnamefont {T.~D.}\ \bibnamefont {Skinner}}, \bibinfo {author}
  {\bibfnamefont {J.}~\bibnamefont {Wunderlich}}, \bibinfo {author}
  {\bibfnamefont {V.}~\bibnamefont {Nov{\'a}k}}, \bibinfo {author}
  {\bibfnamefont {R.~P.}\ \bibnamefont {Campion}}, \bibinfo {author}
  {\bibfnamefont {B.~L.}\ \bibnamefont {Gallagher}}, \bibinfo {author}
  {\bibfnamefont {E.~K.}\ \bibnamefont {Vehstedt}}, \bibinfo {author}
  {\bibfnamefont {L.~P.}\ \bibnamefont {Z{\^a}rbo}}, \bibinfo {author}
  {\bibfnamefont {K.}~\bibnamefont {V{\'y}born{\'y}}}, \bibinfo {author}
  {\bibfnamefont {A.~J.}\ \bibnamefont {Ferguson}}, \ and\ \bibinfo {author}
  {\bibfnamefont {T.}~\bibnamefont {Jungwirth}},\ }\bibfield  {title} {\enquote
  {\bibinfo {title} {{An antidamping spin-orbit torque originating from the
  Berry curvature}},}\ }\href@noop {} {\bibfield  {journal} {\bibinfo
  {journal} {Nat. Nanotechnol.}\ }\textbf {\bibinfo {volume} {9}},\ \bibinfo
  {pages} {211} (\bibinfo {year} {2014})},\ \bibinfo {note}
  {article}\BibitemShut {NoStop}%
\bibitem [{\citenamefont {Hayashi}\ \emph {et~al.}(2014)\citenamefont
  {Hayashi}, \citenamefont {Kim}, \citenamefont {Yamanouchi},\ and\
  \citenamefont {Ohno}}]{Hayashi2014}%
  \BibitemOpen
  \bibfield  {author} {\bibinfo {author} {\bibfnamefont {Masamitsu}\
  \bibnamefont {Hayashi}}, \bibinfo {author} {\bibfnamefont {Junyeon}\
  \bibnamefont {Kim}}, \bibinfo {author} {\bibfnamefont {Michihiko}\
  \bibnamefont {Yamanouchi}}, \ and\ \bibinfo {author} {\bibfnamefont {Hideo}\
  \bibnamefont {Ohno}},\ }\bibfield  {title} {\enquote {\bibinfo {title}
  {{Quantitative characterization of the spin-orbit torque using harmonic Hall
  voltage measurements}},}\ }\href {\doibase 10.1103/PhysRevB.89.144425}
  {\bibfield  {journal} {\bibinfo  {journal} {Phys. Rev. B}\ }\textbf {\bibinfo
  {volume} {89}},\ \bibinfo {pages} {144425} (\bibinfo {year}
  {2014})}\BibitemShut {NoStop}%
\bibitem [{\citenamefont {Yu}\ \emph {et~al.}(2014)\citenamefont {Yu},
  \citenamefont {Upadhyaya}, \citenamefont {Fan}, \citenamefont {Alzate},
  \citenamefont {Jiang}, \citenamefont {Wong}, \citenamefont {Takei},
  \citenamefont {Bender}, \citenamefont {Chang}, \citenamefont {Jiang},
  \citenamefont {Lang}, \citenamefont {Tang}, \citenamefont {Wang},
  \citenamefont {Tserkovnyak}, \citenamefont {Amiri},\ and\ \citenamefont
  {Wang}}]{Yu2014}%
  \BibitemOpen
  \bibfield  {author} {\bibinfo {author} {\bibfnamefont {Guoqiang}\
  \bibnamefont {Yu}}, \bibinfo {author} {\bibfnamefont {Pramey}\ \bibnamefont
  {Upadhyaya}}, \bibinfo {author} {\bibfnamefont {Yabin}\ \bibnamefont {Fan}},
  \bibinfo {author} {\bibfnamefont {Juan~G.}\ \bibnamefont {Alzate}}, \bibinfo
  {author} {\bibfnamefont {Wanjun}\ \bibnamefont {Jiang}}, \bibinfo {author}
  {\bibfnamefont {Kin~L.}\ \bibnamefont {Wong}}, \bibinfo {author}
  {\bibfnamefont {So}~\bibnamefont {Takei}}, \bibinfo {author} {\bibfnamefont
  {Scott~A.}\ \bibnamefont {Bender}}, \bibinfo {author} {\bibfnamefont {Li-Te}\
  \bibnamefont {Chang}}, \bibinfo {author} {\bibfnamefont {Ying}\ \bibnamefont
  {Jiang}}, \bibinfo {author} {\bibfnamefont {Murong}\ \bibnamefont {Lang}},
  \bibinfo {author} {\bibfnamefont {Jianshi}\ \bibnamefont {Tang}}, \bibinfo
  {author} {\bibfnamefont {Yong}\ \bibnamefont {Wang}}, \bibinfo {author}
  {\bibfnamefont {Yaroslav}\ \bibnamefont {Tserkovnyak}}, \bibinfo {author}
  {\bibfnamefont {Pedram~Khalili}\ \bibnamefont {Amiri}}, \ and\ \bibinfo
  {author} {\bibfnamefont {Kang~L.}\ \bibnamefont {Wang}},\ }\bibfield  {title}
  {\enquote {\bibinfo {title} {Switching of perpendicular magnetization by
  spin-orbit torques in the absence of external magnetic fields},}\ }\href@noop
  {} {\bibfield  {journal} {\bibinfo  {journal} {Nat. Nanotechnol.}\ }\textbf
  {\bibinfo {volume} {9}},\ \bibinfo {pages} {548} (\bibinfo {year}
  {2014})}\BibitemShut {NoStop}%
\bibitem [{\citenamefont {Freimuth}\ \emph {et~al.}(2014)\citenamefont
  {Freimuth}, \citenamefont {Bl{\"u}gel},\ and\ \citenamefont
  {Mokrousov}}]{Freimuth2014}%
  \BibitemOpen
  \bibfield  {author} {\bibinfo {author} {\bibfnamefont {Frank}\ \bibnamefont
  {Freimuth}}, \bibinfo {author} {\bibfnamefont {Stefan}\ \bibnamefont
  {Bl{\"u}gel}}, \ and\ \bibinfo {author} {\bibfnamefont {Yuriy}\ \bibnamefont
  {Mokrousov}},\ }\bibfield  {title} {\enquote {\bibinfo {title} {{Spin-orbit
  torques in Co/Pt(111) and Mn/W(001) magnetic bilayers from first
  principles}},}\ }\href {\doibase 10.1103/PhysRevB.90.174423} {\bibfield
  {journal} {\bibinfo  {journal} {Phys. Rev. B}\ }\textbf {\bibinfo {volume}
  {90}},\ \bibinfo {pages} {174423} (\bibinfo {year} {2014})}\BibitemShut
  {NoStop}%
\bibitem [{\citenamefont {Kim}\ \emph {et~al.}(2017)\citenamefont {Kim},
  \citenamefont {Lee}, \citenamefont {Sinova}, \citenamefont {Lee},\ and\
  \citenamefont {Stiles}}]{Kim2017}%
  \BibitemOpen
  \bibfield  {author} {\bibinfo {author} {\bibfnamefont {Kyoung-Whan}\
  \bibnamefont {Kim}}, \bibinfo {author} {\bibfnamefont {Kyung-Jin}\
  \bibnamefont {Lee}}, \bibinfo {author} {\bibfnamefont {Jairo}\ \bibnamefont
  {Sinova}}, \bibinfo {author} {\bibfnamefont {Hyun-Woo}\ \bibnamefont {Lee}},
  \ and\ \bibinfo {author} {\bibfnamefont {M.~D.}\ \bibnamefont {Stiles}},\
  }\bibfield  {title} {\enquote {\bibinfo {title} {Spin-orbit torques from
  interfacial spin-orbit coupling for various interfaces},}\ }\href {\doibase
  10.1103/PhysRevB.96.104438} {\bibfield  {journal} {\bibinfo  {journal} {Phys.
  Rev. B}\ }\textbf {\bibinfo {volume} {96}},\ \bibinfo {pages} {104438}
  (\bibinfo {year} {2017})}\BibitemShut {NoStop}%
\bibitem [{\citenamefont {Mahfouzi}\ and\ \citenamefont
  {Kioussis}(2018)}]{Mahfouzi2018}%
  \BibitemOpen
  \bibfield  {author} {\bibinfo {author} {\bibfnamefont {Farzad}\ \bibnamefont
  {Mahfouzi}}\ and\ \bibinfo {author} {\bibfnamefont {Nicholas}\ \bibnamefont
  {Kioussis}},\ }\bibfield  {title} {\enquote {\bibinfo {title}
  {{First-principles study of the angular dependence of the spin-orbit torque
  in Pt/Co and Pd/Co bilayers}},}\ }\href {\doibase 10.1103/PhysRevB.97.224426}
  {\bibfield  {journal} {\bibinfo  {journal} {Phys. Rev. B}\ }\textbf {\bibinfo
  {volume} {97}},\ \bibinfo {pages} {224426} (\bibinfo {year}
  {2018})}\BibitemShut {NoStop}%
\bibitem [{\citenamefont {Tanaka}\ \emph {et~al.}(2008)\citenamefont {Tanaka},
  \citenamefont {Kontani}, \citenamefont {Naito}, \citenamefont {Naito},
  \citenamefont {Hirashima}, \citenamefont {Yamada},\ and\ \citenamefont
  {Inoue}}]{Tanaka2008}%
  \BibitemOpen
  \bibfield  {author} {\bibinfo {author} {\bibfnamefont {T.}~\bibnamefont
  {Tanaka}}, \bibinfo {author} {\bibfnamefont {H.}~\bibnamefont {Kontani}},
  \bibinfo {author} {\bibfnamefont {M.}~\bibnamefont {Naito}}, \bibinfo
  {author} {\bibfnamefont {T.}~\bibnamefont {Naito}}, \bibinfo {author}
  {\bibfnamefont {D.~S.}\ \bibnamefont {Hirashima}}, \bibinfo {author}
  {\bibfnamefont {K.}~\bibnamefont {Yamada}}, \ and\ \bibinfo {author}
  {\bibfnamefont {J.}~\bibnamefont {Inoue}},\ }\bibfield  {title} {\enquote
  {\bibinfo {title} {{Intrinsic spin Hall effect and orbital Hall effect in
  $4d$ and $5d$ transition metals}},}\ }\href {\doibase
  10.1103/PhysRevB.77.165117} {\bibfield  {journal} {\bibinfo  {journal} {Phys.
  Rev. B}\ }\textbf {\bibinfo {volume} {77}},\ \bibinfo {pages} {165117}
  (\bibinfo {year} {2008})}\BibitemShut {NoStop}%
\bibitem [{\citenamefont {Kontani}\ \emph {et~al.}(2009)\citenamefont
  {Kontani}, \citenamefont {Tanaka}, \citenamefont {Hirashima}, \citenamefont
  {Yamada},\ and\ \citenamefont {Inoue}}]{Kontani2009}%
  \BibitemOpen
  \bibfield  {author} {\bibinfo {author} {\bibfnamefont {H.}~\bibnamefont
  {Kontani}}, \bibinfo {author} {\bibfnamefont {T.}~\bibnamefont {Tanaka}},
  \bibinfo {author} {\bibfnamefont {D.~S.}\ \bibnamefont {Hirashima}}, \bibinfo
  {author} {\bibfnamefont {K.}~\bibnamefont {Yamada}}, \ and\ \bibinfo {author}
  {\bibfnamefont {J.}~\bibnamefont {Inoue}},\ }\bibfield  {title} {\enquote
  {\bibinfo {title} {{Giant Orbital Hall Effect in Transition Metals: Origin of
  Large Spin and Anomalous Hall Effects}},}\ }\href {\doibase
  10.1103/PhysRevLett.102.016601} {\bibfield  {journal} {\bibinfo  {journal}
  {Phys. Rev. Lett.}\ }\textbf {\bibinfo {volume} {102}},\ \bibinfo {pages}
  {016601} (\bibinfo {year} {2009})}\BibitemShut {NoStop}%
\bibitem [{\citenamefont {Go}\ \emph {et~al.}(2018)\citenamefont {Go},
  \citenamefont {Jo}, \citenamefont {Kim},\ and\ \citenamefont {Lee}}]{Go2018}%
  \BibitemOpen
  \bibfield  {author} {\bibinfo {author} {\bibfnamefont {Dongwook}\
  \bibnamefont {Go}}, \bibinfo {author} {\bibfnamefont {Daegeun}\ \bibnamefont
  {Jo}}, \bibinfo {author} {\bibfnamefont {Changyoung}\ \bibnamefont {Kim}}, \
  and\ \bibinfo {author} {\bibfnamefont {Hyun-Woo}\ \bibnamefont {Lee}},\
  }\bibfield  {title} {\enquote {\bibinfo {title} {{Intrinsic Spin and Orbital
  Hall Effects from Orbital Texture}},}\ }\href {\doibase
  10.1103/PhysRevLett.121.086602} {\bibfield  {journal} {\bibinfo  {journal}
  {Phys. Rev. Lett.}\ }\textbf {\bibinfo {volume} {121}},\ \bibinfo {pages}
  {086602} (\bibinfo {year} {2018})}\BibitemShut {NoStop}%
\bibitem [{\citenamefont {Jo}\ \emph {et~al.}(2018)\citenamefont {Jo},
  \citenamefont {Go},\ and\ \citenamefont {Lee}}]{Jo2018}%
  \BibitemOpen
  \bibfield  {author} {\bibinfo {author} {\bibfnamefont {Daegeun}\ \bibnamefont
  {Jo}}, \bibinfo {author} {\bibfnamefont {Dongwook}\ \bibnamefont {Go}}, \
  and\ \bibinfo {author} {\bibfnamefont {Hyun-Woo}\ \bibnamefont {Lee}},\
  }\bibfield  {title} {\enquote {\bibinfo {title} {Gigantic intrinsic orbital
  hall effects in weakly spin-orbit coupled metals},}\ }\href {\doibase
  10.1103/PhysRevB.98.214405} {\bibfield  {journal} {\bibinfo  {journal} {Phys.
  Rev. B}\ }\textbf {\bibinfo {volume} {98}},\ \bibinfo {pages} {214405}
  (\bibinfo {year} {2018})}\BibitemShut {NoStop}%
\bibitem [{Cub()}]{Cubic_d_comment}%
  \BibitemOpen
  \href@noop {} {}\bibinfo {note} {The OHE arises in more realistic $d$ models
  that take into account one of the following complexities; next-nearest
  neighbor hopping, extra orbitals ($s$ or $p$), or non-simple-cubic
  structure.}\BibitemShut {Stop}%
\bibitem [{Sup()}]{Supplementary}%
  \BibitemOpen
  \href@noop {} {}\bibinfo {note} {See Supplementary Material}\BibitemShut
  {NoStop}%
\bibitem [{\citenamefont {Wang}\ and\ \citenamefont
  {Callaway}(1974)}]{Wang1974}%
  \BibitemOpen
  \bibfield  {author} {\bibinfo {author} {\bibfnamefont {C.~S.}\ \bibnamefont
  {Wang}}\ and\ \bibinfo {author} {\bibfnamefont {J.}~\bibnamefont
  {Callaway}},\ }\bibfield  {title} {\enquote {\bibinfo {title} {Band structure
  of nickel: Spin-orbit coupling, the fermi surface, and the optical
  conductivity},}\ }\href {\doibase 10.1103/PhysRevB.9.4897} {\bibfield
  {journal} {\bibinfo  {journal} {Phys. Rev. B}\ }\textbf {\bibinfo {volume}
  {9}},\ \bibinfo {pages} {4897--4907} (\bibinfo {year} {1974})}\BibitemShut
  {NoStop}%
\bibitem [{\citenamefont {Sunko}\ \emph {et~al.}(2017)\citenamefont {Sunko},
  \citenamefont {Rosner}, \citenamefont {Kushwaha}, \citenamefont {Khim},
  \citenamefont {Mazzola}, \citenamefont {Bawden}, \citenamefont {Clark},
  \citenamefont {Riley}, \citenamefont {Kasinathan}, \citenamefont {Haverkort},
  \citenamefont {Kim}, \citenamefont {Hoesch}, \citenamefont {Fujii},
  \citenamefont {Vobornik}, \citenamefont {Mackenzie},\ and\ \citenamefont
  {King}}]{Sunko2017}%
  \BibitemOpen
  \bibfield  {author} {\bibinfo {author} {\bibfnamefont {Veronika}\
  \bibnamefont {Sunko}}, \bibinfo {author} {\bibfnamefont {H.}~\bibnamefont
  {Rosner}}, \bibinfo {author} {\bibfnamefont {P.}~\bibnamefont {Kushwaha}},
  \bibinfo {author} {\bibfnamefont {S.}~\bibnamefont {Khim}}, \bibinfo {author}
  {\bibfnamefont {F.}~\bibnamefont {Mazzola}}, \bibinfo {author} {\bibfnamefont
  {L.}~\bibnamefont {Bawden}}, \bibinfo {author} {\bibfnamefont {O.~J.}\
  \bibnamefont {Clark}}, \bibinfo {author} {\bibfnamefont {J.~M.}\ \bibnamefont
  {Riley}}, \bibinfo {author} {\bibfnamefont {D.}~\bibnamefont {Kasinathan}},
  \bibinfo {author} {\bibfnamefont {M.~W.}\ \bibnamefont {Haverkort}}, \bibinfo
  {author} {\bibfnamefont {T.~K.}\ \bibnamefont {Kim}}, \bibinfo {author}
  {\bibfnamefont {M.}~\bibnamefont {Hoesch}}, \bibinfo {author} {\bibfnamefont
  {J.}~\bibnamefont {Fujii}}, \bibinfo {author} {\bibfnamefont
  {I.}~\bibnamefont {Vobornik}}, \bibinfo {author} {\bibfnamefont {A.~P.}\
  \bibnamefont {Mackenzie}}, \ and\ \bibinfo {author} {\bibfnamefont
  {P.~D.~C.}\ \bibnamefont {King}},\ }\bibfield  {title} {\enquote {\bibinfo
  {title} {Maximal rashba-like spin splitting via kinetic-energy-coupled
  inversion-symmetry breaking},}\ }\href {https://doi.org/10.1038/nature23898}
  {\bibfield  {journal} {\bibinfo  {journal} {Nature}\ }\textbf {\bibinfo
  {volume} {549}},\ \bibinfo {pages} {492} (\bibinfo {year}
  {2017})}\BibitemShut {NoStop}%
\bibitem [{\citenamefont {Freimuth}\ \emph {et~al.}(2010)\citenamefont
  {Freimuth}, \citenamefont {Bl{\"u}gel},\ and\ \citenamefont
  {Mokrousov}}]{Freimuth2010}%
  \BibitemOpen
  \bibfield  {author} {\bibinfo {author} {\bibfnamefont {Frank}\ \bibnamefont
  {Freimuth}}, \bibinfo {author} {\bibfnamefont {Stefan}\ \bibnamefont
  {Bl{\"u}gel}}, \ and\ \bibinfo {author} {\bibfnamefont {Yuriy}\ \bibnamefont
  {Mokrousov}},\ }\bibfield  {title} {\enquote {\bibinfo {title} {{Anisotropic
  Spin Hall Effect from First Principles}},}\ }\href {\doibase
  10.1103/PhysRevLett.105.246602} {\bibfield  {journal} {\bibinfo  {journal}
  {Phys. Rev. Lett.}\ }\textbf {\bibinfo {volume} {105}},\ \bibinfo {pages}
  {246602} (\bibinfo {year} {2010})}\BibitemShut {NoStop}%
\bibitem [{\citenamefont {Zhu}\ \emph {et~al.}(2018)\citenamefont {Zhu},
  \citenamefont {Ralph},\ and\ \citenamefont {Buhrman}}]{Zhu2018}%
  \BibitemOpen
  \bibfield  {author} {\bibinfo {author} {\bibfnamefont {Lijun}\ \bibnamefont
  {Zhu}}, \bibinfo {author} {\bibfnamefont {Daniel.~C.}\ \bibnamefont {Ralph}},
  \ and\ \bibinfo {author} {\bibfnamefont {Robert~A.}\ \bibnamefont
  {Buhrman}},\ }\bibfield  {title} {\enquote {\bibinfo {title} {Highly
  {E}fficient {S}pin-{C}urrent {G}eneration by the {S}pin {H}all {E}ffect in
  {A}u$_{1\ensuremath{-}x}${P}t$_{x}$},}\ }\href {\doibase
  10.1103/PhysRevApplied.10.031001} {\bibfield  {journal} {\bibinfo  {journal}
  {Phys. Rev. Appl.}\ }\textbf {\bibinfo {volume} {10}},\ \bibinfo {pages}
  {031001} (\bibinfo {year} {2018})}\BibitemShut {NoStop}%
\bibitem [{\citenamefont {Du}\ \emph {et~al.}(2018)\citenamefont {Du},
  \citenamefont {Karube}, \citenamefont {Gamou}, \citenamefont {Ryu},
  \citenamefont {Takahashi}, \citenamefont {Kohda},\ and\ \citenamefont
  {Nitta}}]{Du2018}%
  \BibitemOpen
  \bibfield  {author} {\bibinfo {author} {\bibfnamefont {Ye}~\bibnamefont
  {Du}}, \bibinfo {author} {\bibfnamefont {Shutaro}\ \bibnamefont {Karube}},
  \bibinfo {author} {\bibfnamefont {Hiromu}\ \bibnamefont {Gamou}}, \bibinfo
  {author} {\bibfnamefont {Jeongchun}\ \bibnamefont {Ryu}}, \bibinfo {author}
  {\bibfnamefont {Saburo}\ \bibnamefont {Takahashi}}, \bibinfo {author}
  {\bibfnamefont {Makoto}\ \bibnamefont {Kohda}}, \ and\ \bibinfo {author}
  {\bibfnamefont {Junsaku}\ \bibnamefont {Nitta}},\ }\href@noop {} {\enquote
  {\bibinfo {title} {Anomalous spin orbit torques with large rashba spin orbit
  coupling in epitaxial {P}t/{C}o bilayers},}\ } (\bibinfo {year} {2018}),\
  \Eprint {http://arxiv.org/abs/arXiv:1807.10867} {arXiv:1807.10867}
  \BibitemShut {NoStop}%
\bibitem [{FM_()}]{FM_SHE_comment}%
  \BibitemOpen
  \href@noop {} {}\bibinfo {note} {Increasing the SOC in the FM may result in
  sizable SHE in the FM, as pointed out in Refs. \cite{Amin2019,Wang2019}.
  However, the SHE in the FM leads to an antisymmetric accumulation of the spin
  within the FM layer, leading to a non-uniform tilting of the magnetization.
  Thus, when the thickness of the FM is smaller than the spin diffusion length,
  this effect is expected to vanish. On the other hand, the OT mechanism
  proposed in this Letter is not affected by the thickness of the FM, and
  results in uniform tilting of the magnetization.}\BibitemShut {Stop}%
\bibitem [{ato()}]{atomic_ordering_comment}%
  \BibitemOpen
  \href@noop {} {}\bibinfo {note} {Note that the atomic ordering persists
  locally even in polycrsytalline samples. However, at the interface
  mixing(demixing) nature of the NM and FM elements can suppress(enhance) the
  crystallinity at the interface. On the other hand, in the bulk, the
  crystallinity dependence of the OT will be weaker.}\BibitemShut {Stop}%
\bibitem [{\citenamefont {Du}\ \emph {et~al.}(2014)\citenamefont {Du},
  \citenamefont {Wang}, \citenamefont {Yang},\ and\ \citenamefont
  {Hammel}}]{Du2014}%
  \BibitemOpen
  \bibfield  {author} {\bibinfo {author} {\bibfnamefont {Chunhui}\ \bibnamefont
  {Du}}, \bibinfo {author} {\bibfnamefont {Hailong}\ \bibnamefont {Wang}},
  \bibinfo {author} {\bibfnamefont {Fengyuan}\ \bibnamefont {Yang}}, \ and\
  \bibinfo {author} {\bibfnamefont {P.~Chris}\ \bibnamefont {Hammel}},\
  }\bibfield  {title} {\enquote {\bibinfo {title} {Systematic variation of
  spin-orbit coupling with $d$-orbital filling: Large inverse spin hall effect
  in $3d$ transition metals},}\ }\href {\doibase 10.1103/PhysRevB.90.140407}
  {\bibfield  {journal} {\bibinfo  {journal} {Phys. Rev. B}\ }\textbf {\bibinfo
  {volume} {90}},\ \bibinfo {pages} {140407(R)} (\bibinfo {year}
  {2014})}\BibitemShut {NoStop}%
\bibitem [{\citenamefont {Qu}\ \emph {et~al.}(2015)\citenamefont {Qu},
  \citenamefont {Huang},\ and\ \citenamefont {Chien}}]{Qu2015}%
  \BibitemOpen
  \bibfield  {author} {\bibinfo {author} {\bibfnamefont {D.}~\bibnamefont
  {Qu}}, \bibinfo {author} {\bibfnamefont {S.~Y.}\ \bibnamefont {Huang}}, \
  and\ \bibinfo {author} {\bibfnamefont {C.~L.}\ \bibnamefont {Chien}},\
  }\bibfield  {title} {\enquote {\bibinfo {title} {Inverse spin hall effect in
  cr: Independence of antiferromagnetic ordering},}\ }\href {\doibase
  10.1103/PhysRevB.92.020418} {\bibfield  {journal} {\bibinfo  {journal} {Phys.
  Rev. B}\ }\textbf {\bibinfo {volume} {92}},\ \bibinfo {pages} {020418(R)}
  (\bibinfo {year} {2015})}\BibitemShut {NoStop}%
\bibitem [{\citenamefont {Miao}\ \emph {et~al.}(2013)\citenamefont {Miao},
  \citenamefont {Huang}, \citenamefont {Qu},\ and\ \citenamefont
  {Chien}}]{Miao2013}%
  \BibitemOpen
  \bibfield  {author} {\bibinfo {author} {\bibfnamefont {B.~F.}\ \bibnamefont
  {Miao}}, \bibinfo {author} {\bibfnamefont {S.~Y.}\ \bibnamefont {Huang}},
  \bibinfo {author} {\bibfnamefont {D.}~\bibnamefont {Qu}}, \ and\ \bibinfo
  {author} {\bibfnamefont {C.~L.}\ \bibnamefont {Chien}},\ }\bibfield  {title}
  {\enquote {\bibinfo {title} {{Inverse Spin Hall Effect in a Ferromagnetic
  Metal}},}\ }\href {\doibase 10.1103/PhysRevLett.111.066602} {\bibfield
  {journal} {\bibinfo  {journal} {Phys. Rev. Lett.}\ }\textbf {\bibinfo
  {volume} {111}},\ \bibinfo {pages} {066602} (\bibinfo {year}
  {2013})}\BibitemShut {NoStop}%
\bibitem [{\citenamefont {Tsukahara}\ \emph {et~al.}(2014)\citenamefont
  {Tsukahara}, \citenamefont {Ando}, \citenamefont {Kitamura}, \citenamefont
  {Emoto}, \citenamefont {Shikoh}, \citenamefont {Delmo}, \citenamefont
  {Shinjo},\ and\ \citenamefont {Shiraishi}}]{Tsukahara2014}%
  \BibitemOpen
  \bibfield  {author} {\bibinfo {author} {\bibfnamefont {Ayaka}\ \bibnamefont
  {Tsukahara}}, \bibinfo {author} {\bibfnamefont {Yuichiro}\ \bibnamefont
  {Ando}}, \bibinfo {author} {\bibfnamefont {Yuta}\ \bibnamefont {Kitamura}},
  \bibinfo {author} {\bibfnamefont {Hiroyuki}\ \bibnamefont {Emoto}}, \bibinfo
  {author} {\bibfnamefont {Eiji}\ \bibnamefont {Shikoh}}, \bibinfo {author}
  {\bibfnamefont {Michael~P.}\ \bibnamefont {Delmo}}, \bibinfo {author}
  {\bibfnamefont {Teruya}\ \bibnamefont {Shinjo}}, \ and\ \bibinfo {author}
  {\bibfnamefont {Masashi}\ \bibnamefont {Shiraishi}},\ }\bibfield  {title}
  {\enquote {\bibinfo {title} {Self-induced inverse spin hall effect in
  permalloy at room temperature},}\ }\href {\doibase
  10.1103/PhysRevB.89.235317} {\bibfield  {journal} {\bibinfo  {journal} {Phys.
  Rev. B}\ }\textbf {\bibinfo {volume} {89}},\ \bibinfo {pages} {235317}
  (\bibinfo {year} {2014})}\BibitemShut {NoStop}%
\bibitem [{\citenamefont {Go}\ \emph {et~al.}(2017)\citenamefont {Go},
  \citenamefont {Hanke}, \citenamefont {Buhl}, \citenamefont {Freimuth},
  \citenamefont {Bihlmayer}, \citenamefont {Lee}, \citenamefont {Mokrousov},\
  and\ \citenamefont {Bl{\"u}gel}}]{Go2017}%
  \BibitemOpen
  \bibfield  {author} {\bibinfo {author} {\bibfnamefont {Dongwook}\
  \bibnamefont {Go}}, \bibinfo {author} {\bibfnamefont {Jan-Philipp}\
  \bibnamefont {Hanke}}, \bibinfo {author} {\bibfnamefont {Patrick~M.}\
  \bibnamefont {Buhl}}, \bibinfo {author} {\bibfnamefont {Frank}\ \bibnamefont
  {Freimuth}}, \bibinfo {author} {\bibfnamefont {Gustav}\ \bibnamefont
  {Bihlmayer}}, \bibinfo {author} {\bibfnamefont {Hyun-Woo}\ \bibnamefont
  {Lee}}, \bibinfo {author} {\bibfnamefont {Yuriy}\ \bibnamefont {Mokrousov}},
  \ and\ \bibinfo {author} {\bibfnamefont {Stefan}\ \bibnamefont
  {Bl{\"u}gel}},\ }\bibfield  {title} {\enquote {\bibinfo {title} {Toward
  surface orbitronics: giant orbital magnetism from the orbital rashba effect
  at the surface of sp-metals},}\ }\href {http://dx.doi.org/10.1038/srep46742}
  {\bibfield  {journal} {\bibinfo  {journal} {Sci. Rep.}\ }\textbf {\bibinfo
  {volume} {7}},\ \bibinfo {pages} {46742} (\bibinfo {year}
  {2017})}\BibitemShut {NoStop}%
\bibitem [{\citenamefont {Chen}\ \emph {et~al.}(2018)\citenamefont {Chen},
  \citenamefont {Liu}, \citenamefont {Yang}, \citenamefont {Shi}, \citenamefont
  {Hu}, \citenamefont {Li},\ and\ \citenamefont {Zeng}}]{Chen2018}%
  \BibitemOpen
  \bibfield  {author} {\bibinfo {author} {\bibfnamefont {Xi}~\bibnamefont
  {Chen}}, \bibinfo {author} {\bibfnamefont {Yang}\ \bibnamefont {Liu}},
  \bibinfo {author} {\bibfnamefont {Guang}\ \bibnamefont {Yang}}, \bibinfo
  {author} {\bibfnamefont {Hui}\ \bibnamefont {Shi}}, \bibinfo {author}
  {\bibfnamefont {Chen}\ \bibnamefont {Hu}}, \bibinfo {author} {\bibfnamefont
  {Minghua}\ \bibnamefont {Li}}, \ and\ \bibinfo {author} {\bibfnamefont
  {Haibo}\ \bibnamefont {Zeng}},\ }\bibfield  {title} {\enquote {\bibinfo
  {title} {{Giant antidamping orbital torque originating from the orbital
  Rashba-Edelstein effect in ferromagnetic heterostructures}},}\ }\href
  {\doibase 10.1038/s41467-018-05057-z} {\bibfield  {journal} {\bibinfo
  {journal} {Nat. Commun.}\ }\textbf {\bibinfo {volume} {9}},\ \bibinfo {pages}
  {2569} (\bibinfo {year} {2018})}\BibitemShut {NoStop}%
\bibitem [{\citenamefont {Emori}\ \emph {et~al.}(2016)\citenamefont {Emori},
  \citenamefont {Nan}, \citenamefont {Belkessam}, \citenamefont {Wang},
  \citenamefont {Matyushov}, \citenamefont {Babroski}, \citenamefont {Gao},
  \citenamefont {Lin},\ and\ \citenamefont {Sun}}]{Emori2016}%
  \BibitemOpen
  \bibfield  {author} {\bibinfo {author} {\bibfnamefont {Satoru}\ \bibnamefont
  {Emori}}, \bibinfo {author} {\bibfnamefont {Tianxiang}\ \bibnamefont {Nan}},
  \bibinfo {author} {\bibfnamefont {Amine~M.}\ \bibnamefont {Belkessam}},
  \bibinfo {author} {\bibfnamefont {Xinjun}\ \bibnamefont {Wang}}, \bibinfo
  {author} {\bibfnamefont {Alexei~D.}\ \bibnamefont {Matyushov}}, \bibinfo
  {author} {\bibfnamefont {Christopher~J.}\ \bibnamefont {Babroski}}, \bibinfo
  {author} {\bibfnamefont {Yuan}\ \bibnamefont {Gao}}, \bibinfo {author}
  {\bibfnamefont {Hwaider}\ \bibnamefont {Lin}}, \ and\ \bibinfo {author}
  {\bibfnamefont {Nian~X.}\ \bibnamefont {Sun}},\ }\bibfield  {title} {\enquote
  {\bibinfo {title} {Interfacial spin-orbit torque without bulk spin-orbit
  coupling},}\ }\href {\doibase 10.1103/PhysRevB.93.180402} {\bibfield
  {journal} {\bibinfo  {journal} {Phys. Rev. B}\ }\textbf {\bibinfo {volume}
  {93}},\ \bibinfo {pages} {180402(R)} (\bibinfo {year} {2016})}\BibitemShut
  {NoStop}%
\bibitem [{\citenamefont {Amin}\ \emph {et~al.}(2019)\citenamefont {Amin},
  \citenamefont {Li}, \citenamefont {Stiles},\ and\ \citenamefont
  {Haney}}]{Amin2019}%
  \BibitemOpen
  \bibfield  {author} {\bibinfo {author} {\bibfnamefont {V.~P.}\ \bibnamefont
  {Amin}}, \bibinfo {author} {\bibfnamefont {Junwen}\ \bibnamefont {Li}},
  \bibinfo {author} {\bibfnamefont {M.~D.}\ \bibnamefont {Stiles}}, \ and\
  \bibinfo {author} {\bibfnamefont {P.~M.}\ \bibnamefont {Haney}},\ }\bibfield
  {title} {\enquote {\bibinfo {title} {Intrinsic spin currents in
  ferromagnets},}\ }\href {\doibase 10.1103/PhysRevB.99.220405} {\bibfield
  {journal} {\bibinfo  {journal} {Phys. Rev. B}\ }\textbf {\bibinfo {volume}
  {99}},\ \bibinfo {pages} {220405} (\bibinfo {year} {2019})}\BibitemShut
  {NoStop}%
\bibitem [{\citenamefont {{Wang}}\ \emph {et~al.}(2019)\citenamefont {{Wang}},
  \citenamefont {{Wang}}, \citenamefont {{Amin}}, \citenamefont {{Wang}},
  \citenamefont {{radhakrishnan}}, \citenamefont {{Davidson}}, \citenamefont
  {{Allen}}, \citenamefont {{Silva}}, \citenamefont {{Ohldag}}, \citenamefont
  {{Balzar}}, \citenamefont {{Zink}}, \citenamefont {{Haney}}, \citenamefont
  {{Xiao}}, \citenamefont {{Cahill}}, \citenamefont {{Lorenz}},\ and\
  \citenamefont {{Fan}}}]{Wang2019}%
  \BibitemOpen
  \bibfield  {author} {\bibinfo {author} {\bibfnamefont {W.}~\bibnamefont
  {{Wang}}}, \bibinfo {author} {\bibfnamefont {T.}~\bibnamefont {{Wang}}},
  \bibinfo {author} {\bibfnamefont {V.~P.}\ \bibnamefont {{Amin}}}, \bibinfo
  {author} {\bibfnamefont {Y.}~\bibnamefont {{Wang}}}, \bibinfo {author}
  {\bibfnamefont {A.}~\bibnamefont {{radhakrishnan}}}, \bibinfo {author}
  {\bibfnamefont {A.}~\bibnamefont {{Davidson}}}, \bibinfo {author}
  {\bibfnamefont {S.~R.}\ \bibnamefont {{Allen}}}, \bibinfo {author}
  {\bibfnamefont {T.~J.}\ \bibnamefont {{Silva}}}, \bibinfo {author}
  {\bibfnamefont {H.}~\bibnamefont {{Ohldag}}}, \bibinfo {author}
  {\bibfnamefont {D.}~\bibnamefont {{Balzar}}}, \bibinfo {author}
  {\bibfnamefont {B.~L.}\ \bibnamefont {{Zink}}}, \bibinfo {author}
  {\bibfnamefont {P.~M.}\ \bibnamefont {{Haney}}}, \bibinfo {author}
  {\bibfnamefont {J.~Q.}\ \bibnamefont {{Xiao}}}, \bibinfo {author}
  {\bibfnamefont {D.~G.}\ \bibnamefont {{Cahill}}}, \bibinfo {author}
  {\bibfnamefont {V.~O.}\ \bibnamefont {{Lorenz}}}, \ and\ \bibinfo {author}
  {\bibfnamefont {X.}~\bibnamefont {{Fan}}},\ }\href@noop {} {\enquote
  {\bibinfo {title} {Anomalous spin-orbit torques in magnetic single-layer
  films},}\ } (\bibinfo {year} {2019}),\ \Eprint
  {http://arxiv.org/abs/arXiv:1902.05490} {arXiv:1902.05490} \BibitemShut
  {NoStop}%
\end{thebibliography}%


\begin{thebibliography}{1}%
\makeatletter
\providecommand \@ifxundefined [1]{%
 \@ifx{#1\undefined}
}%
\providecommand \@ifnum [1]{%
 \ifnum #1\expandafter \@firstoftwo
 \else \expandafter \@secondoftwo
 \fi
}%
\providecommand \@ifx [1]{%
 \ifx #1\expandafter \@firstoftwo
 \else \expandafter \@secondoftwo
 \fi
}%
\providecommand \natexlab [1]{#1}%
\providecommand \enquote  [1]{``#1''}%
\providecommand \bibnamefont  [1]{#1}%
\providecommand \bibfnamefont [1]{#1}%
\providecommand \citenamefont [1]{#1}%
\providecommand \href@noop [0]{\@secondoftwo}%
\providecommand \href [0]{\begingroup \@sanitize@url \@href}%
\providecommand \@href[1]{\@@startlink{#1}\@@href}%
\providecommand \@@href[1]{\endgroup#1\@@endlink}%
\providecommand \@sanitize@url [0]{\catcode `\\12\catcode `\$12\catcode
  `\&12\catcode `\#12\catcode `\^12\catcode `\_12\catcode `\%12\relax}%
\providecommand \@@startlink[1]{}%
\providecommand \@@endlink[0]{}%
\providecommand \url  [0]{\begingroup\@sanitize@url \@url }%
\providecommand \@url [1]{\endgroup\@href {#1}{\urlprefix }}%
\providecommand \urlprefix  [0]{URL }%
\providecommand \Eprint [0]{\href }%
\providecommand \doibase [0]{http://dx.doi.org/}%
\providecommand \selectlanguage [0]{\@gobble}%
\providecommand \bibinfo  [0]{\@secondoftwo}%
\providecommand \bibfield  [0]{\@secondoftwo}%
\providecommand \translation [1]{[#1]}%
\providecommand \BibitemOpen [0]{}%
\providecommand \bibitemStop [0]{}%
\providecommand \bibitemNoStop [0]{.\EOS\space}%
\providecommand \EOS [0]{\spacefactor3000\relax}%
\providecommand \BibitemShut  [1]{\csname bibitem#1\endcsname}%
\let\auto@bib@innerbib\@empty
%</preamble>
\bibitem [{\citenamefont {Go}\ \emph {et~al.}(2018)\citenamefont {Go},
  \citenamefont {Jo}, \citenamefont {Kim},\ and\ \citenamefont {Lee}}]{Go2018_supp}%
  \BibitemOpen
  \bibfield  {author} {\bibinfo {author} {\bibfnamefont {D.}~\bibnamefont
  {Go}}, \bibinfo {author} {\bibfnamefont {D.}~\bibnamefont {Jo}}, \bibinfo
  {author} {\bibfnamefont {C.}~\bibnamefont {Kim}}, \ and\ \bibinfo {author}
  {\bibfnamefont {H.-W.}\ \bibnamefont {Lee}},\ }\href {\doibase
  10.1103/PhysRevLett.121.086602} {\bibfield  {journal} {\bibinfo  {journal}
  {Phys. Rev. Lett.}\ }\textbf {\bibinfo {volume} {121}},\ \bibinfo {pages}
  {086602} (\bibinfo {year} {2018})}\BibitemShut {NoStop}%
\end{thebibliography}

\clearpage

\newpage

\pagebreak

\widetext

\let\addcontentsline\oldaddcontentsline% Restore \addcontentsline

\setcounter{equation}{0}
\setcounter{figure}{0}
\setcounter{table}{0}
\setcounter{page}{1}

\renewcommand{\theequation}{S\arabic{equation}}
\renewcommand{\thefigure}{S\arabic{figure}}
\renewcommand{\bibnumfmt}[1]{[S#1]}
\renewcommand{\citenumfont}[1]{S#1}
\renewcommand{\thepage}{S\arabic{page}}  

\setlength{\textfloatsep}{25pt}
\setlength{\parskip}{0pt}

\begin{center}
	\textbf{\large \bf Supplementary Material for
\\
``Orbital Torque: Torque Generation by Orbital Current Injection''}
\end{center}
\begin{center}
	{ Dongwook Go and Hyun-Woo Lee$^*$}
\end{center}

\tableofcontents

\subsection{A. \ Tight-Binding Model}
\label{sec:tight-binding_model}

The tight-binding model for a magnetic bilayer presented in the Letter is composed of a nonmagnet (NM) and a ferromagnet (FM). The numbers of layers for the NM and the FM are $N_\mathrm{NM}$ and $N_\mathrm{FM}$, respectively. We assume the simple cubic structure for both NM and FM with only nearest neighbor hoppings allowed. We also assume that the layer is periodic in $x$ and $y$ directions, and the layers are stacked along $z$ direction. Thus, the NM is located from $z=1$ to $z=N_\mathrm{NM}$ and the FM is located from $z = N_\mathrm{NM} + 1$ to $z=N_\mathrm{NM} + N_\mathrm{FM}$ (in unit of the lattice spacing $a$), and we use the Bloch theorem for $x$ and $y$ directions by introducing the crystal momentum $\mathbf{k}=(k_x, k_y)$. The total Hamiltonian is formally written as
\begin{eqnarray}
H_\mathrm{tot}(\mathbf{k})
=
\left(
\begin{array}{ccccc|ccccc}
H_\mathrm{NM}^\mathrm{2d} (\mathbf{k}) & T_\mathrm{NM}^\dagger  & \cdots &  0 & 0 & 0 & 0 & \cdots & 0 & 0 \\
T_\mathrm{NM} & H_\mathrm{NM}^\mathrm{2d} (\mathbf{k}) & \cdots & 0 & 0 & 0 & 0 & \cdots & 0 & 0 \\
\vdots & \vdots & \ddots & \vdots & \vdots & \vdots & \vdots & \ddots & \vdots & \vdots \\
0 & 0 & \cdots & H_\mathrm{NM}^\mathrm{2d} (\mathbf{k}) & T_\mathrm{NM}^\dagger & 0 & 0 & \cdots & 0 & 0 \\
0 & 0 & \cdots & T_\mathrm{NM} & H_\mathrm{NM}^\mathrm{2d} (\mathbf{k}) & T_\mathrm{int}^\dagger & 0 & \cdots & 0 & 0 \\
\hline
0 & 0 & \cdots & 0 & T_\mathrm{int} & H_\mathrm{FM}^\mathrm{2d} (\mathbf{k}) & T_\mathrm{FM}\dagger & \cdots & 0 & 0 \\
0 & 0 & \cdots & 0 & 0 & T_\mathrm{FM} & H_\mathrm{FM}^\mathrm{2d} (\mathbf{k}) & \cdots & 0 & 0 \\
\vdots & \vdots & \ddots & \vdots & \vdots & \vdots & \vdots & \ddots & \vdots & \vdots \\
0 & 0 & \cdots & 0 & 0 & 0 & 0 & \cdots & H_\mathrm{FM}^\mathrm{2d} (\mathbf{k}) & T_\mathrm{FM}^\dagger \\
0 & 0 & \cdots & 0 & 0 & 0 & 0 & \cdots & T_\mathrm{FM} & H_\mathrm{FM}^\mathrm{2d} (\mathbf{k})
\end{array}
\right),
\end{eqnarray}
where $H_\mathrm{NM(FM)}^\mathrm{2d}(\mathbf{k})$ is the Hamiltonian for a two-dimensional NM(FM) layer, $T_\mathrm{NM(FM)}$ is the hopping between nearest NM(FM) layers, and $T_\mathrm{int}$ is the interface hopping between the last NM layer ($z=N_\mathrm{NM}$) and the first FM layer ($z=N_\mathrm{NM}+1$).

\subsubsection{1. \  NM}

We assume the NM hosts $sp_\alpha\ (\alpha=x,y,z)$ orbitals at each site, which was introduced in Ref.~\cite{Go2018_supp}. Writing the Hamiltonian in a finite film structure is straightforward as follows. The Hamiltonian within each two-dimensional NM layer consists of the kinetic energy and spin-orbit coupling (SOC) parts:
\begin{eqnarray}
H_\mathrm{NM}^\mathrm{2d} (\mathbf{k})
=
H_\mathrm{NM}^\mathrm{kin} (\mathbf{k})
+
H_\mathrm{NM}^\mathrm{soc} (\mathbf{k}).
\end{eqnarray}
First, the kinetic part is 
\begin{eqnarray}
H_\mathrm{NM}^\mathrm{kin}(\mathbf{k})
=
\left(
\begin{array}{cccc}
E_s (\mathbf{k}) & 2i\gamma_{sp} \sin (k_x a) & 2i\gamma_{sp} \sin(k_y a) & 0 \\
-2i\gamma_{sp}\sin (k_x a) & E_{p_x} (\mathbf{k}) & 0 & 0 \\ 
-2i\gamma_{sp}\sin (k_y a) & 0 & E_{p_y} (\mathbf{k}) & 0 \\
0 & 0 & 0 & E_{p_z} (\mathbf{k})
\end{array}
\right)
\otimes
\mathrm{I}_\mathrm{2\times 2},
\end{eqnarray}
where 
\begin{subequations}
\begin{eqnarray}
E_s (\mathbf{k}) 
&=&
E_s
-2 t_s 
\left[ 
\cos (k_x a) + \cos (k_y a)
\right],
\\
E_{p_x} (\mathbf{k}) 
&=&
E_{p_x}
+ 2 t_{p\sigma} \cos (k_x a)
-2 t_{p\pi}
\cos (k_y a),
\\
E_{p_y} (\mathbf{k}) 
&=& 
E_{p_y}
- 2 t_{p\pi} \cos (k_x a)
+2 t_{p\sigma}
\cos (k_y a),
\\
E_{p_z} (\mathbf{k})
&=&
E_{p_z}
- 2 t_{p\pi} 
\left[
\cos (k_x a) + \cos (k_y a)
\right],
\end{eqnarray}
\end{subequations}
and $\mathrm{I}_\mathrm{2\times 2}$ is an identity operator in the spin space. Here, the basis states are 
\begin{eqnarray}
\label{eq:Wannier}
\ket{\varphi^{(z)}_{l\sigma\mathbf{k}}} 
=
\sum_{\mathbf{R}}
e^{i\mathbf{k}\cdot\mathbf{R}}
\ket{\phi^{(z)}_{l\sigma\mathbf{R}}},
\end{eqnarray}
where $\ket{\phi^{(z)}_{l\sigma\mathbf{R}}}$ is a Wannier function localized at the Bravais lattice $\mathbf{R}=(R_x, R_y)$ with its orbital character $l=s, p_x, p_y, p_z$ and spin $\sigma$, which is defined in a layer located at $z$. For the Wannier states, $E_s$, $E_{p_\alpha}$ are onsite energies for $s$ and $p_\alpha$ orbitals, and $t_s$, $t_{p\sigma(\pi)}$, $\gamma_{sp}$ are the nearest hopping amplitudes between $s$ orbitals, between $p$ orbitals via $\sigma(\pi)$ bonding, and between $s$ and $p$ orbitals, respectively. Second, the SOC part is 
\begin{eqnarray}
H_\mathrm{NM}^\mathrm{so} = \frac{\alpha_\mathrm{so}^\mathrm{NM}}{\hbar^2} \mathbf{L}^{(p)} \cdot \mathbf{S},
\end{eqnarray}
where $\mathbf{S}$ is the spin operator and $\mathbf{L}^{(p)}$ is the orbital angular momentum (OAM) operator in $p$ orbital space. Here, $\alpha_\mathrm{so}^\mathrm{NM}>0$ is the strength of the SOC in the NM. The OAM operator is explicitly expressed in a matrix representation
\begin{eqnarray}
L_x^{(p)}
=
\hbar
\begin{pmatrix}
0 & 0 & 0  \\
0 & 0 & -i \\
0 & i & 0
\end{pmatrix},
\ \ \
L_y^{(p)}
=
\hbar
\begin{pmatrix}
0 & 0 & i \\
0 & 0 & 0 \\
-i & 0 & 0 
\end{pmatrix},
\ \ \
L_z^{(p)}
=
\hbar
\begin{pmatrix}
0 & -i & 0 \\
i & 0 & 0 \\
0 & 0 & 0 
\end{pmatrix},
\end{eqnarray}
with $p_x$, $p_y$, and $p_z$ orbital Wannier functions. Finally, the interlayer coupling between neighboring NM layers is described as
\begin{eqnarray}
T_\mathrm{NM}
=
\left(
\begin{array}{cccc}
-t_{ss} & 0 & 0 & -\gamma_{sp} \\
0 & -t_{p\pi} & 0 & 0 \\
0 & 0 & -t_{p\pi} & 0 \\
\gamma_{sp} & 0 & 0 & t_{p\sigma}
\end{array}
\right)
\otimes
\mathrm{I}_\mathrm{2\times 2},
\end{eqnarray}
where the basis states for the row and column are $\bra{\varphi^{(z+1)}_{l\sigma\mathbf{k}}}$ and $\ket{\varphi^{(z)}_{l'\sigma'\mathbf{k}}}$, respectively, for $z=1, \cdots, N_\mathrm{NM}-1$.

\subsubsection{2. \ FM}

In the FM, we assume there are $d_\beta\ (\beta = xy, yz, zx, x^2-y^2, z^2)$ orbitals at each site. The Hamiltonian within each two-dimensional layer is 
\begin{eqnarray}
H_\mathrm{FM}^\mathrm{(2d)} (\mathbf{k}) = H_\mathrm{FM}^\mathrm{kin} (\mathbf{k}) + H_\mathrm{FM}^\mathrm{so} + H_\mathrm{FM}^\mathrm{xc},
\end{eqnarray}
where each term describes kinetic energy, SOC, and exchange interaction with magnetization, respectively. The kinetic energy term is
\begin{eqnarray}
H_\mathrm{FM}^\mathrm{kin(2d)} (\mathbf{k})
=
\left(
\begin{array}{ccccc}
E_{d_{xy}} (\mathbf{k}) & 0 & 0 & 0 & 0  \\
0 & E_{d_{yz}} (\mathbf{k}) & 0 & 0 & 0 \\
0 & 0 &E_{d_{zx}} (\mathbf{k}) & 0 & 0 \\
0 & 0 & 0 &  E_{d_{x^2-y^2}} (\mathbf{k}) & 0 \\
0 & 0 & 0& 0 & E_{d_{z^2}} (\mathbf{k})
\end{array}
\right)
\otimes
\mathrm{I}_\mathrm{2\times 2}
,
\end{eqnarray}
where
\begin{subequations}
\begin{eqnarray}
E_{d_{xy}} (\mathbf{k})
&=&
E_{d_{xy}}
+2 t_{d\pi} \cos (k_x a) + 2 t_{d\pi} \cos(k_y a),
\\
E_{d_{yz}} (\mathbf{k})
&=&
E_{d_{yz}}
-2 t_{d\delta} \cos (k_x a) + 2 t_{d\pi} \cos(k_y a),
\\
E_{d_{zx}} (\mathbf{k})
&=&
E_{d_{zx}}
+2 t_{d\pi} \cos (k_x a) - 2 t_{d\delta} \cos(k_y a),
\\
E_{d_{x^2-y^2}} (\mathbf{k})
&=&
E_{d_{x^2-y^2}}
- [(3/2) t_{d\sigma} + (1/2) t_{d\delta}] [ \cos(k_x a) + \cos (k_y a) ],
\\
E_{d_{z^2}} (\mathbf{k})
&=&
E_{d_{z^2}}
- [(1/2) t_{d\sigma} + (3/2) t_{d\delta}] [ \cos(k_x a) + \cos (k_y a) ].
\end{eqnarray}
\end{subequations}
Here, $E_{d_\beta}$ is the onsite energy of the $d_\beta$ orbital, and $t_{d\sigma}$, $t_{d\pi}$, $t_{d\delta}$ are nearest neighbor hoppings between $d$ orbitals via $\sigma$, $\pi$, $\delta$ bondings, respectively. The basis states are defined similarly as Eq.~(\ref{eq:Wannier}) but for $d_\beta$ orbital Wannier functions. The SOC term is 
\begin{eqnarray}
\label{eq:SOC_FM}
H_\mathrm{FM}^\mathrm{so} = \frac{\alpha_\mathrm{so}^\mathrm{FM}}{\hbar^2} \mathbf{L}^{(d)} \cdot \mathbf{S},
\end{eqnarray}
where $\alpha_\mathrm{so}^\mathrm{FM} > 0$ is the SOC strength. Here, $\mathbf{L}^{(d)}$ is the OAM operator in $d$ orbital space, whose matrix representation is written as 
\begin{eqnarray}
L_x^{(d)} = 
\hbar
\left(
\begin{array}{ccccc}
0 & 0 & -i & -i & -\sqrt{3} i \\
0 & 0 & 0 & 0 & 0 \\
i & 0 & 0 & 0 & 0 \\
i & 0 & 0 & 0 & 0 \\
\sqrt{3}i & 0 & 0 & 0 & 0 
\end{array}
\right),
\ \ \
L_y^{(d)} =
\hbar
\left(
\begin{array}{ccccc}
0 & i & 0 & 0 & 0 \\
-i & 0 & 0 & 0 & 0 \\
0 & 0 & 0 & -i & \sqrt{3}i \\
0 & 0 & i & 0 & 0 \\
0 & 0 & -\sqrt{3}i & 0 & 0 
\end{array}
\right),
\ \ \ 
L_z^{(d)} =
\hbar
\left(
\begin{array}{ccccc}
0 & 0 & 0 & 2i & 0 \\
0 & 0 & i & 0 & 0 \\
0 & -i & 0 & 0 & 0 \\
-2i & 0 & 0 & 0 & 0 \\
0 & 0 & 0 & 0 & 0 
\end{array}
\right),
\nonumber
\\
\end{eqnarray}
where the basis states are $d_{xy}$, $d_{yz}$, $d_{zx}$, $d_{x^2-y^2}$, $d_{z^2}$ orbital Wannier functions. The exchange interaction is 
\begin{eqnarray}
\label{eq:exchange}
H_\mathrm{FM}^\mathrm{xc} = \frac{J}{\hbar} \hat{\mathbf{M}}\cdot\mathbf{S},
\end{eqnarray}
where $J >0$ is the strength of the exchange interaction, and $\hat{\mathbf{M}}$ is the direction of the magnetization. We assume $\hat{\mathbf{M}}=\hat{\mathbf{z}}$ in the calculation. The interlayer coupling between neighboring FM layers is
\begin{eqnarray}
T_\mathrm{FM}
=
\left(
\begin{array}{ccccc}
-t_{d\delta} & 0 & 0 & 0 & 0 \\
0 & t_{d\pi} & 0 & 0 & 0 \\
0 & 0 & t_{d\pi} & 0 & 0 \\
0 & 0 & 0 & -t_{d\delta} & 0 \\
0 & 0 & 0 & 0 & -t_{d\sigma}
\end{array}
\right)
\otimes
\mathrm{I}_\mathrm{2\times 2}
,
\end{eqnarray}
where the basis for the row and column are $\bra{\varphi^{(z+1)}_{l\sigma \mathbf{k}}}$ and $\ket{\varphi^{(z)}_{l'\sigma'\mathbf{k}}}$, respectively, for $z=N_\mathrm{NM}+1, \cdots, N_\mathrm{NM}+N_\mathrm{FM}-1$.  

\subsubsection{3. \ Interface}

At the interface, there are hoppings between the last NM layer ($z=N_\mathrm{NM}$) and the first FM layer ($z=N_\mathrm{NM}+1$), which are expressed in 
\begin{eqnarray}
T_\mathrm{interface}
=
\left(
\begin{array}{cccc}
0 & 0 & \gamma_{pd\pi} & 0 \\
0 & 0 & \gamma_{pd\pi} & 0 \\
0 & 0 & 0 & 0 \\
0 & 0 & 0 & 0 \\
0 & 0 & 0 & -\gamma_{pd\sigma}
\end{array}
\right)
\otimes
\mathrm{I}_\mathrm{2\times 2}
, 
\end{eqnarray} 
where the basis for the row and column are $\bra{\varphi^{(N_\mathrm{NM}+1)}_{l\sigma\mathbf{k}}}$ and $\ket{\varphi^{(N_\mathrm{NM})}_{l'\sigma'\mathbf{k}}}$, respectively. Here, $\gamma_{pd\sigma(\pi)}$ is the nearest neighbor hopping between $p$ and $d$ orbitals via $\sigma(\pi)$ hoppings. We neglect the hopping from a $s$ orbital in the NM to $d$ orbitals in the FM, since the $s$ orbital does not carry the OAM, thus not affecting the orbital injection. 

\subsubsection{4. \ Parameter Setting}

For the tight-binding model defined above, parameters which we used for the calculation in Figs. 2 and 3 of the Letter are set as
\begin{eqnarray}
\label{eq:param_NM}
E_{s} = 3.2, \
E_{p_x} = E_{p_y} = E_{p_z} = -0.5, \ 
t_{s} = 0.5, \
t_{p\sigma} = 0.5, \
t_{p\pi} = 0.2, \
\gamma_{sp} = 0.5, \
\alpha_\mathrm{so}^\mathrm{NM} = 0, \ 
\end{eqnarray}
for the NM, 
\begin{eqnarray}
\label{eq:param_FM}
E_{d_{xy}} = E_{d_{yz}} = E_{d_{zx}} = E_{d_{x^2-y^2}} = E_{d_{z^2}} = -0.5, \
t_{d\sigma} = 0.1, \
t_{d\pi} = 0.05, \
t_{d\delta} = 0.02, \
J = 0.5, \
\alpha_\mathrm{so}^\mathrm{FM} = 0.1 
\end{eqnarray}
for the FM, and
\begin{eqnarray}
\label{eq:param_interface}
\gamma_{pd\sigma} = 0.4, \
\gamma_{pd\pi} = 0.1
\end{eqnarray}
for the interface. All parameters are expressed in unit of $\mathrm{eV}$.

\subsection{B. \ Spatial Profiles of the Orbital and Spin Hall Currents}

%========================================================
\begin{figure}[t!]
\includegraphics[angle=0, width=0.55\textwidth]{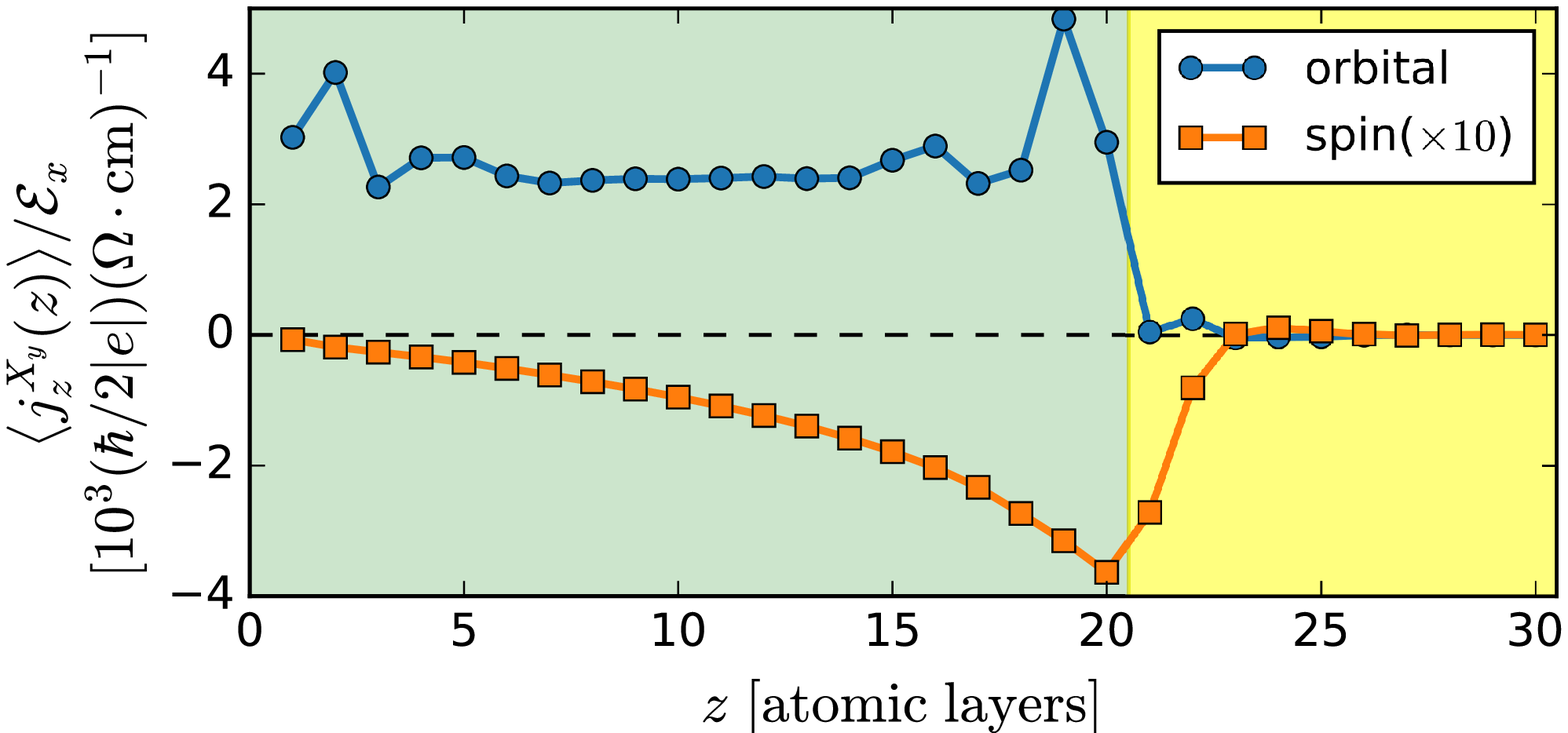}
\caption{
\label{fig:current_spatial_profile}
Spatial profiles of the orbital (blue circles) and spin (orange squares, 10x enlarged) Hall currents in the NM region ($1 \leq z \leq 20$) and the FM region ($21 \leq z \leq 30$). The Fermi energy is set to $E_\mathrm{F}=-0.9\ \mathrm{eV}$. 
}
\end{figure}
%========================================================

Linear responses of the orbital and spin Hall currents are evaluated using the Kubo formula as follows:
\begin{eqnarray}
\label{eq:Kubo_current}
\left\langle \delta j_z^{X_y} (z) \right\rangle
=
-e\hbar \mathcal{E}_x
\sum_{nm}
\int \frac{d^2k}{(2\pi)^2}
(f_{n\mathbf{k}} - f_{m\mathbf{k}})
\textup{Im}
\left[
\frac{
	\bra{u_{n\mathbf{k}}} j_z^{X_y}(z) \ket{u_{m\mathbf{k}}}
	\bra{u_{m\mathbf{k}}} v_x \ket{u_{n\mathbf{k}}}}
{(E_{n\mathbf{k}} - E_{m\mathbf{k}} + i\Gamma)^2}
\right],
\end{eqnarray}
where 
\begin{eqnarray}
j_z^{X_y}(z) 
=
\frac{1}{2}
\sum_{z'}
\left[
P(z')
\frac{1}{2}
\left\{
v_z, X_y
\right\}
P(z)
+
P(z)
\frac{1}{2}
\left\{
v_z, X_y
\right\}
P(z')
\right]
\end{eqnarray}
is the spin/orbital ($\mathbf{X}=\mathbf{L}$ or $\mathbf{S}$) current operator defined in a layer at $z$. Here, $P(z)$ is the projection operator to a layer at $z$. The velocity operator along the $z$ direction is defined as
\begin{eqnarray}
v_z = \frac{1}{i\hbar} \sum_{zz'}
(z-z')
P(z)
H
P(z').
\end{eqnarray}
Figure~\ref{fig:current_spatial_profile} shows spatial profiles of the orbital and spin Hall currents obtained from the tight-binding model introduced in Sec. \ref{sec:tight-binding_model}. The parameters are set as Eqs. \eqref{eq:param_NM}-\eqref{eq:param_interface}, and the numbers of the NM and FM layers are $N_\mathrm{NM}=20$ and $N_\mathrm{FM}=10$. For this calculation, we set the Fermi energy as $E_\mathrm{F}=-0.9\ \mathrm{eV}$. We find that the orbital Hall conductivity in the NM region is more than $\approx 2,000\ (\hbar/2|e|)(\Omega\mathrm{cm})^{-1}$. In the FM region, part of the orbital Hall current is injected, which is converted to the spin current by the SOC of the FM [Eq~\eqref{eq:SOC_FM}]. We also find the spin current in the NM region, which is decaying from the interface. This is because reflected current from the interface becomes spin-polarized. The decay is due to finite spectral broadening $\Gamma= 25\ \mathrm{meV}$ in Eq.~\eqref{eq:Kubo_current}.

\subsection{C. \ Correlation between $\left\langle X_y \right\rangle^\mathrm{FM}$ and $\left\langle X_x \right\rangle^\mathrm{FM}$ $(X=L,S)$: Fermi Energy Dependence}

%========================================================
\begin{figure}[t!]
\includegraphics[angle=0, width=0.5\textwidth]{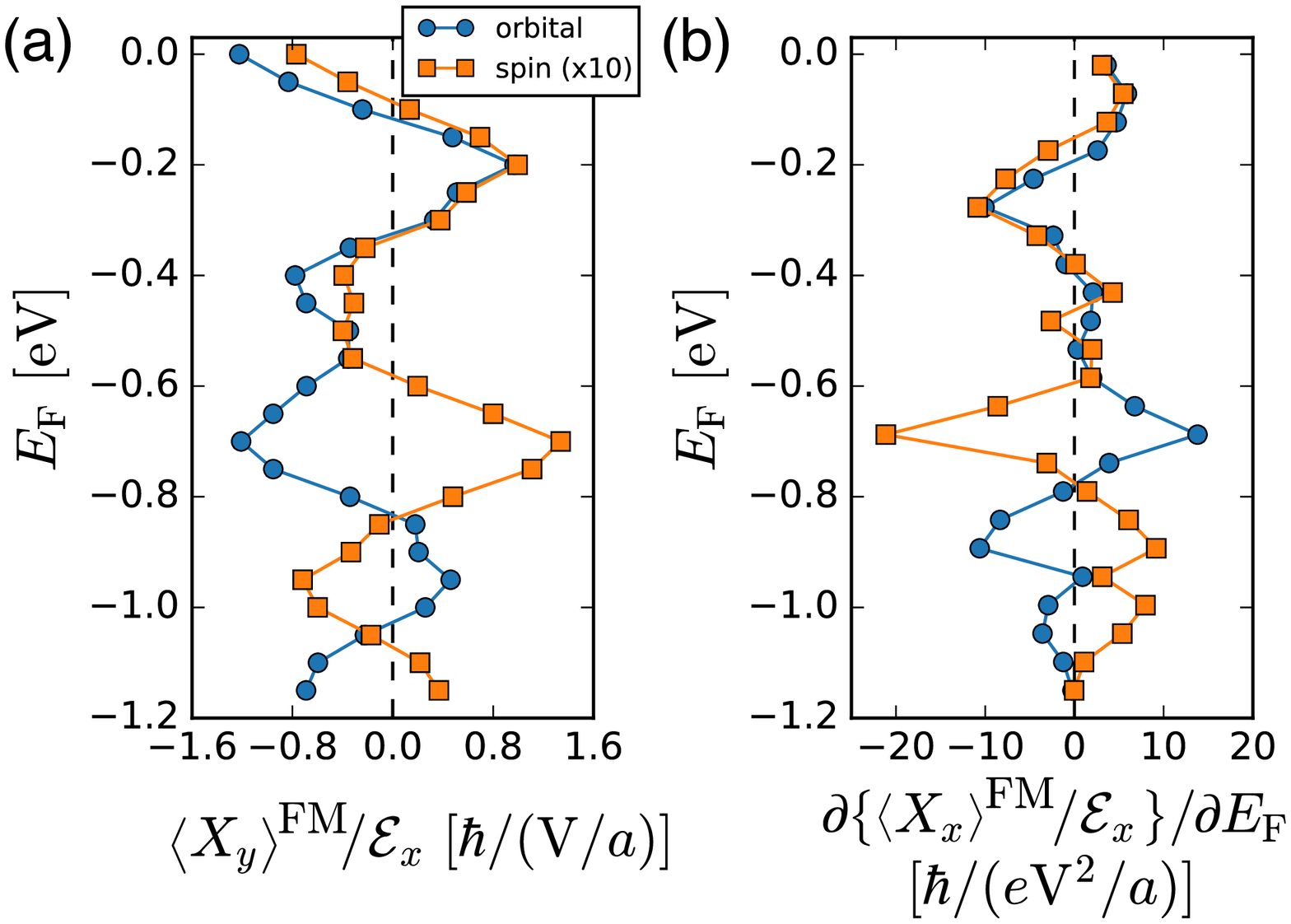}
\caption{
\label{fig:Fermi_dep}
Fermi energy ($E_\mathrm{F}$) dependences of (a) $\left\langle X_y \right\rangle^\mathrm{FM}/\mathcal{E}_x$ and (b) $\partial \{\left\langle X_x \right\rangle^\mathrm{FM}/\mathcal{E}_x \}/\partial E_\mathrm{F}$. For this calculation, the numbers of each layers are set to $N_\mathrm{NM}=8$ and $N_\mathrm{FM}=2$.
}
\end{figure}
%========================================================

Figure \ref{fig:Fermi_dep}(a) shows Fermi energy dependence of $\left\langle X_y\right\rangle^\mathrm{FM}/\mathcal{E}_x$ for $N_\mathrm{NM}=8$ and $N_\mathrm{FM}=2$, where $X=L$ or $S$. This contribution is even under the sign reversal of $\hat{\mathbf{M}}$, and arises from the intraband contribution [Eq.~(2a) of the Letter]. While sign and magnitude vary depending on $E_\mathrm{F}$, the signs of $\left\langle L_y \right\rangle^\mathrm{FM}$ and $\left\langle S_y \right\rangle^\mathrm{FM}$ are opposite in the lower energy range ($-1.2 \lesssim E_\mathrm{F} \lesssim -0.6$) and same in the upper energy range ($-0.5 \lesssim E_\mathrm{F} \lesssim 0$). The $E_\mathrm{F}$-dependence of the sign of the relative ratio between $\left\langle L_y \right\rangle^\mathrm{FM}$ and $\left\langle S_y \right\rangle^\mathrm{FM}$ strikingly resembles the energy dependence of the spin-orbit correlation $\left\langle \mathbf{L}\cdot\mathbf{S}\right\rangle_\mathrm{eq}^\mathrm{FM}$ in Fig.~2(b).

However, the variation of $\left\langle X_y \right\rangle^\mathrm{FM}/\mathcal{E}_x$ [Fig.~\ref{fig:Fermi_dep}(a)] differs from the variation of $\left\langle X_x \right\rangle^\mathrm{FM}/\mathcal{E}_x$ [Fig.~4(a)]. A reason for such difference is due to the fact that while $\left\langle X_y \right\rangle^\mathrm{FM}$ arises from the {\it intraband} contribution for the states at the Fermi {\it surface} [Eq.~2(a)] $\left\langle X_x \right\rangle^\mathrm{FM}$ arises from the {\it interband} contribution for the states in the Fermi {\it sea} [Eq.~2(b)]. However, for each state in the band structure, they have strong correlations. To demonstrate this point, we present in Fig.~\ref{fig:Fermi_dep}(b)  a plot of $\partial \{\left\langle X_x \right\rangle^\mathrm{FM}/\mathcal{E}_x \}/\partial E_\mathrm{F}$, which corresponds to the contribution within the energy slice near $E_\mathrm{F}$. By comparing this with $\left\langle X_y \right\rangle^\mathrm{FM}/\mathcal{E}_x$ [Fig.~\ref{fig:Fermi_dep}(a)], we find strong resemblance for both orbital and spin over the whole range of $E_\mathrm{F}$, except that their relative signs are opposite.

\subsection{D. \ Spin-Orbit Coupling Dependence}

\begin{figure}[t!]
\includegraphics[angle=0, width=1.0\textwidth]{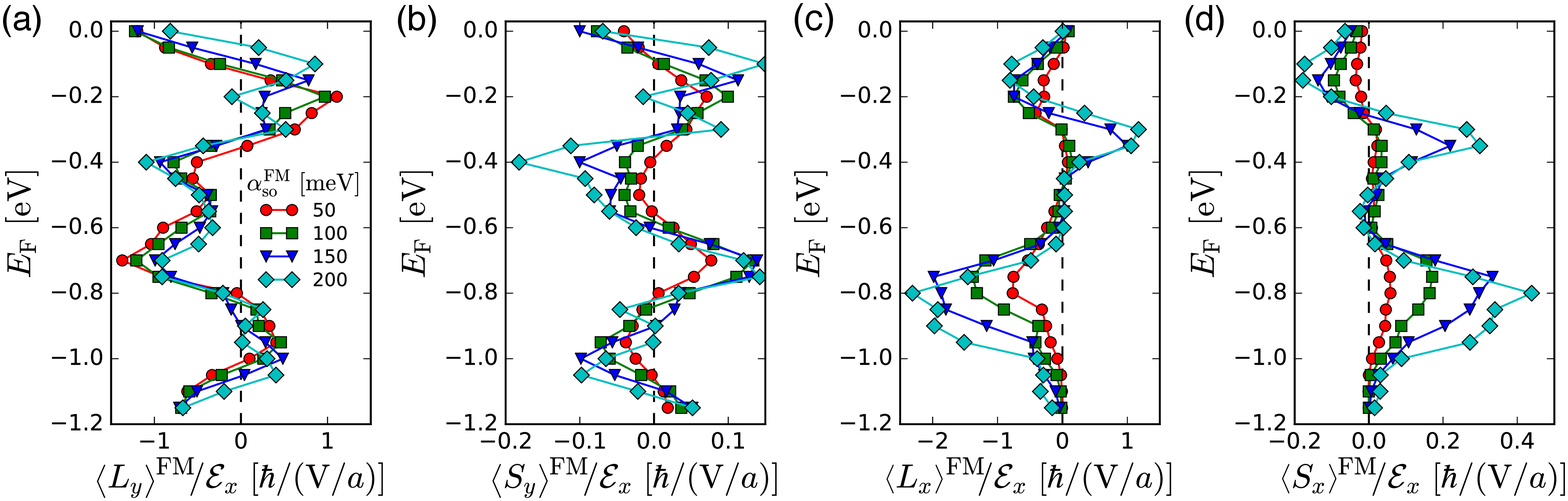}
\caption{
\label{fig:SOC_dep} $E_\mathrm{F}$-dependences of (a) $\left\langle L_y \right\rangle^\mathrm{FM}/\mathcal{E}_x$, (b) $\left\langle S_y \right\rangle^\mathrm{FM}/\mathcal{E}_x$, (c) $\left\langle L_x \right\rangle^\mathrm{FM}/\mathcal{E}_x$, and (d) $\left\langle S_x \right\rangle^\mathrm{FM}/\mathcal{E}_x$ for different values of $\alpha_\mathrm{so}^\mathrm{FM}$. For this calculation, the numbers of each layers are set to $N_\mathrm{NM}=8$ and $N_\mathrm{FM}=2$.}
\end{figure}

Figure \ref{fig:SOC_dep} shows Fermi energy dependences of (a) $\left\langle L_y \right\rangle^\mathrm{FM}/\mathcal{E}_x$, (b) $\left\langle S_y \right\rangle^\mathrm{FM}/\mathcal{E}_x$, (c) $\left\langle L_x \right\rangle^\mathrm{FM}/\mathcal{E}_x$, and (d) $\left\langle S_x \right\rangle^\mathrm{FM}/\mathcal{E}_x$ for different values of $\alpha_\mathrm{so}^\mathrm{FM}$. First, $\left\langle L_y \right\rangle^\mathrm{FM}/\mathcal{E}_x$ remains almost invariant under the increase of $\alpha_\mathrm{so}^\mathrm{FM}$ [Fig.~\ref{fig:SOC_dep}(a)]. This is expected because $\left\langle L_y \right\rangle^\mathrm{FM}$, which results from the orbital Hall effect (OHE) in the NM, is not affected by the SOC of the FM. On the other hand, the rest three quantities exhibit monotonic increase as $\alpha_\mathrm{so}^\mathrm{FM}$ becomes larger within the range $\alpha_\mathrm{so}^\mathrm{FM}\leq 200\ \mathrm{meV}$ [Figs.~\ref{fig:SOC_dep}(b), \ref{fig:SOC_dep}(c), and \ref{fig:SOC_dep}(d)]. The monotonic increase of $\left\langle S_y \right\rangle^\mathrm{FM}/\mathcal{E}_x$ [Fig.~\ref{fig:SOC_dep}(b)] is understandable because $\left\langle S_y \right\rangle^\mathrm{FM}$ is directly converted from the $\left\langle L_y \right\rangle^\mathrm{FM}$ by the SOC in the FM. Also, since $\left\langle S_x \right\rangle^\mathrm{FM}$ results from the precession of $\left\langle S_y \right\rangle^\mathrm{FM}$ by the exchange interaction in the FM [Eq. \eqref{eq:exchange}], monotonic increase of the $\left\langle S_x \right\rangle^\mathrm{FM}/\mathcal{E}_x$ with $\alpha_\mathrm{so}^\mathrm{FM}$ follows that of $\left\langle S_y \right\rangle^\mathrm{FM}/\mathcal{E}_x$ [Fig.~\ref{fig:SOC_dep}(d)]. On the other hand, since the precession of the spin is coupled to the orbital by the SOC in the FM, 
$\left\langle L_x \right\rangle^\mathrm{FM}/\mathcal{E}_x$ is proportional to $\left\langle S_x \right\rangle^\mathrm{FM}/\mathcal{E}_x$. Thus, $\left\langle L_x \right\rangle^\mathrm{FM}/\mathcal{E}_x$ also increases monotonically with increasing $\alpha_\mathrm{so}^\mathrm{FM}$ [Fig.~\ref{fig:SOC_dep}(c)]. 

%========================================================
%========================================================

In Fig.~4(b), the SOC dependence of $\left\langle \tau_d \right\rangle^\mathrm{FM}/\mathcal{E}_x$, which is proportional to $\left\langle S_x \right\rangle^\mathrm{FM}/\mathcal{E}_x$, is shown for fixed Fermi energies $E_\mathrm{F}=-0.15\ \mathrm{eV}$ and $E_\mathrm{F}=-0.80\ \mathrm{eV}$. In Fig.~\ref{fig:SOC_dep}(d), we find that $\left\langle S_x \right\rangle^\mathrm{FM}/\mathcal{E}_x$ monotonically increases with $\alpha_\mathrm{so}^\mathrm{FM}$ at $E_\mathrm{F}=-0.15\ \mathrm{eV}$ and $E_\mathrm{F}=-0.80\ \mathrm{eV}$. Although the magnitude of the peak monotonically increases, the peak position may change at some Fermi energy, i.e. at $E_\mathrm{F}=-0.75\ \mathrm{eV}$. At such Fermi energy, $\left\langle S_x \right\rangle^\mathrm{FM}/\mathcal{E}_x$ may exhibit nonmonotonic behavior. This is due to modification of the band structure with increasing $\alpha_\mathrm{so}^\mathrm{FM}$.

\subsection{E. \ Orbital Torque versus Spin Torque}

\subsubsection{1. \ NM SOC included}

%========================================================
\begin{figure}[b!]
\includegraphics[angle=0, width=0.95\textwidth]{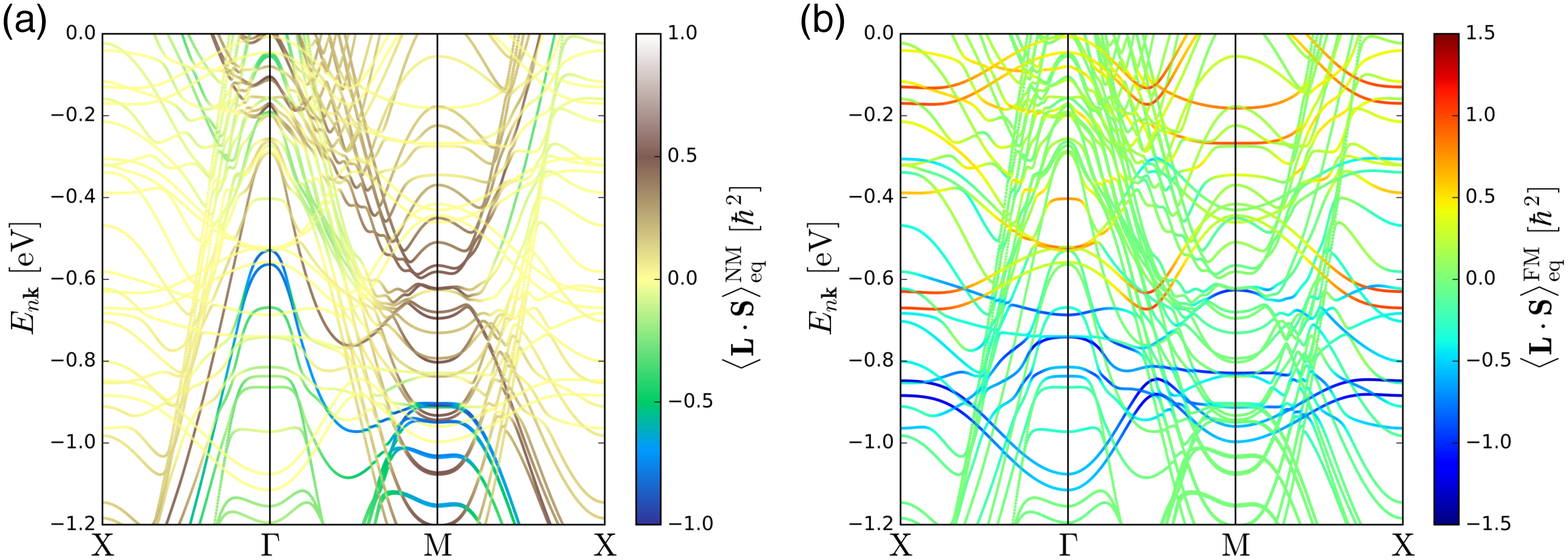}
\caption{
\label{fig:band_soc_NM_FM}
Band structure and the spin-orbit correlations in the (a) NM and (b) FM regions, when the SOC included in both regions. For this calculation, we assumed $N_\mathrm{NM}=8$ and $N_\mathrm{FM}=2$.  
}
\end{figure}
%========================================================

When the SOC in the NM is nonzero, the spin Hall effect (SHE) follows the OHE \cite{Go2018_supp} thus orbital torque (OT) and spin torque (ST) coexist. Relative sign of the OT and ST is determined by the spin-orbit correlations $\left\langle \mathbf{L}\cdot\mathbf{S}\right\rangle_\mathrm{eq}^\mathrm{NM}$ in the NM and $\left\langle \mathbf{L}\cdot\mathbf{S}\right\rangle_\mathrm{eq}^\mathrm{FM}$ in the FM as summarized in Table I of the Letter. In this section, we demonstrate this point from the numerical calculation by setting finite SOC strength in the NM:
\begin{eqnarray}
\label{eq:SOC_NM}
\alpha_\mathrm{so}^\mathrm{NM}=0.2\ \mathrm{eV}.
\end{eqnarray}
Except for $\alpha_\mathrm{so}^\mathrm{NM}$, the rest of the parameters are set equal to Eqs. \eqref{eq:param_NM}, \eqref{eq:param_FM}, and \eqref{eq:param_interface}. Figure~\ref{fig:band_soc_NM_FM} displays the band structure shown in lines and (a) $\left\langle \mathbf{L}\cdot\mathbf{S} \right\rangle_\mathrm{eq}^\mathrm{NM}$ and (b) $\left\langle \mathbf{L}\cdot\mathbf{S} \right\rangle_\mathrm{eq}^\mathrm{FM}$ shown in colors. In the NM, $\left\langle \mathbf{L}\cdot\mathbf{S} \right\rangle_\mathrm{eq}^\mathrm{NM}$ is positive in the upper energy range ($-0.5\ \mathrm{eV} \lesssim E_{n\mathbf{k}} \lesssim 0.0 \ \mathrm{eV}$ near the $\Gamma$-point and $-0.8\ \mathrm{eV} \lesssim E_{n\mathbf{k}} \lesssim 0.0 \ \mathrm{eV}$ near the $\mathrm{M}$-point) and negative in the lower energy range ($-1.2\ \mathrm{eV} \lesssim E_{n\mathbf{k}} \lesssim -0.6 \ \mathrm{eV}$ near the $\Gamma$-point and $-1.2\ \mathrm{eV} \lesssim E_{n\mathbf{k}} \lesssim -0.9 \ \mathrm{eV}$ near the $\mathrm{M}$-point) in general [Fig.~\ref{fig:band_soc_NM_FM}(a)]. Note that this is essentially the same as Fig.~2(b) of Ref.~\cite{Go2018_supp} because all the parameters in the NM regions are set equal to those used in Ref.~\cite{Go2018_supp}. Meanwhile, $\left\langle \mathbf{L}\cdot\mathbf{S} \right\rangle_\mathrm{eq}^\mathrm{FM}$ in Fig.~\ref{fig:band_soc_NM_FM}(b) is almost the same as Fig.~2(b) of the Letter.

%========================================================
\begin{figure}[t!]
\includegraphics[angle=0, width=0.5\textwidth]{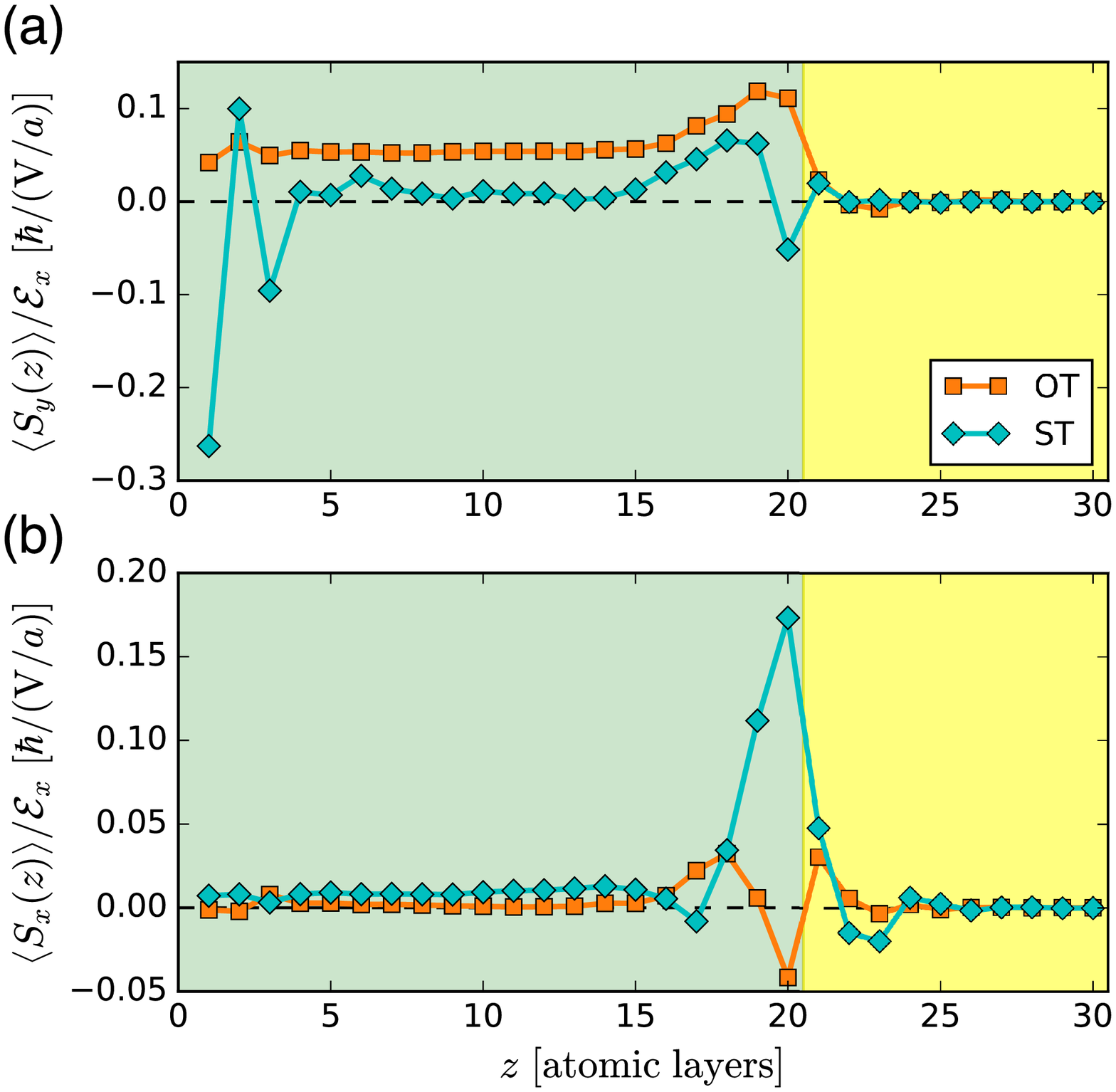}
\caption{
\label{fig:OT_ST}
Spatial profiles of (a) $\left\langle S_y \right\rangle/\mathcal{E}_x$ and (b) $\left\langle S_x \right\rangle/\mathcal{E}_x$ in the NM ($1\leq z \leq 20$) and FM ($21\leq z \leq 30$) regions. The OT and ST contributions are represented by orange squares and cyan diamonds, respectively. The SOC is included in the NM by Eq.~\eqref{eq:SOC_NM} and the corresponding band structure is shown in Fig.~\ref{fig:band_soc_NM_FM}. The Fermi energy is set equal to $E_\mathrm{F}=-0.9\ \mathrm{eV}$. 
}
\end{figure}
%========================================================

In order to differentiate the OT and ST contributions to $\left\langle \mathbf{S}(z) \right\rangle$, we calculate electric field response of $\left\langle \mathbf{S} (z) \right\rangle$ [Eq.~(2) of the Letter] by setting SOC strength parameters in the NM and FM by
(i) $\alpha_\mathrm{so}^\mathrm{NM}=0.2\ \mathrm{eV}$, $\alpha_\mathrm{so}^\mathrm{FM}=0.1\ \mathrm{eV}$, and 
(ii) $\alpha_\mathrm{so}^\mathrm{NM}=0.2\ \mathrm{eV}$, $\alpha_\mathrm{so}^\mathrm{FM}=-0.1\ \mathrm{eV}$. Note that the sign of $\alpha_\mathrm{so}^\mathrm{FM}$ is reversed in (ii). This reversal is motivated by the fact that it reverses the sign of the OT while it does not affect the sign of the ST. By this way, the band structure is barely affected and also the sign of the $\left\langle \mathbf{L}\cdot\mathbf{S} \right\rangle_\mathrm{eq}^\mathrm{FM}$ for the case (ii) becomes opposite to that for the case (i). Thus, we define the OT and ST contributions as
\begin{subequations}
\begin{eqnarray}
\left\langle \mathbf{S} (z) \right\rangle_\mathrm{OT}
&=&
\frac{1}{2}
\left[
\left\langle \mathbf{S} (z) \right\rangle_\mathrm{(i)}
-
\left\langle \mathbf{S} (z) \right\rangle_\mathrm{(ii)}
\right],
\\
\left\langle \mathbf{S} (z) \right\rangle_\mathrm{ST}
&=&
\frac{1}{2}
\left[
\left\langle \mathbf{S} (z) \right\rangle_\mathrm{(i)}
+
\left\langle \mathbf{S} (z) \right\rangle_\mathrm{(ii)}
\right],
\end{eqnarray} 
\label{eq:OT_ST}
\end{subequations}
such that $\left\langle \mathbf{S} (z) \right\rangle_\mathrm{(i)} = \left\langle \mathbf{S} (z) \right\rangle_\mathrm{OT} + \left\langle \mathbf{S} (z) \right\rangle_\mathrm{ST}$. Therefore, we calculate $\left\langle \mathbf{S} (z) \right\rangle_\mathrm{(i)}$ and $\left\langle \mathbf{S} (z) \right\rangle_\mathrm{(i)}$ from the Kubo formula in Eq.~(2) of the Letter, and extract the OT and ST contributions by Eq.~\eqref{eq:OT_ST} for $E_\mathrm{F}=-0.9\ \mathrm{eV}$, where $\left\langle \mathbf{L}\cdot\mathbf{S} \right\rangle_\mathrm{eq}^\mathrm{NM}<0$ and $\left\langle \mathbf{L}\cdot\mathbf{S} \right\rangle_\mathrm{eq}^\mathrm{FM}<0$ [Fig.~\ref{fig:band_soc_NM_FM}]. Thus, the OT and ST contributions are expected to have the same sign.

In Fig. \ref{fig:OT_ST}(a), spatial profile of $\left\langle S_y (z) \right\rangle/\mathcal{E}_x$ is shown in the NM region ($1 \leq z \leq 20$) and the FM region ($21 \leq z \leq 30$). The OT contribution (orange squares) is similar to the result of $\left\langle S_y (z) \right\rangle/\mathcal{E}_x$ when $\alpha_\mathrm{so}^\mathrm{NM}=0$ [Fig.~3(a) of the Letter]. On the other hand, the ST contribution (cyan diamonds) exhibits a standard behavior of the SHE; the spin is accumulated at the boundary. Near $z=1$, sign of $\left\langle S_y (z) \right\rangle/\mathcal{E}_x$ is negative, which implies that the OHE and SHE occurs in the opposite directions. In the absence of the FM, $\left\langle S_y (z) \right\rangle/\mathcal{E}_x$ is positive near $z=20$ (not sown). However, due to presence of the FM attached, $\left\langle S_y (z) \right\rangle/\mathcal{E}_x$ is reduced near $z=20$, which is injected to the FM. We find that the signs of the OT and ST contributions are same in the FM region. This is expected from the spin-orbit correlations of states near $E=-0.9\ \mathrm{eV}$ and from the fact that $\left\langle S_y (z) \right\rangle/\mathcal{E}_x$ results from the intraband contribution at the Fermi surface [Eq.~(2a)]. When $\left\langle S_y \right\rangle$ is injected to the FM, it precesses along the magnetization by the exchange interaction, regardless of whether it is the OT or ST contribution. In the Kubo formula calculation, $\left\langle S_x (z) \right\rangle/\mathcal{E}_x$ is captured by the interband contribution in the Fermi sea [Eq. (2b) of the Letter]. Nevertheless, we find that the signs of the OT and ST contributions are same in the FM region [Fig.~\ref{fig:OT_ST}(b)], which is because {\it all the states} below $E_\mathrm{F}=-0.9\ \mathrm{eV}$ satisfy $\left\langle \mathbf{L}\cdot\mathbf{S} \right\rangle_\mathrm{eq}^\mathrm{NM}<0$ and $\left\langle \mathbf{L}\cdot\mathbf{S} \right\rangle_\mathrm{eq}^\mathrm{FM}<0$.

\subsubsection{2. \ NM SOC included, FM onsite changed}

%========================================================
\begin{figure}[t!]
\includegraphics[angle=0, width=0.95\textwidth]{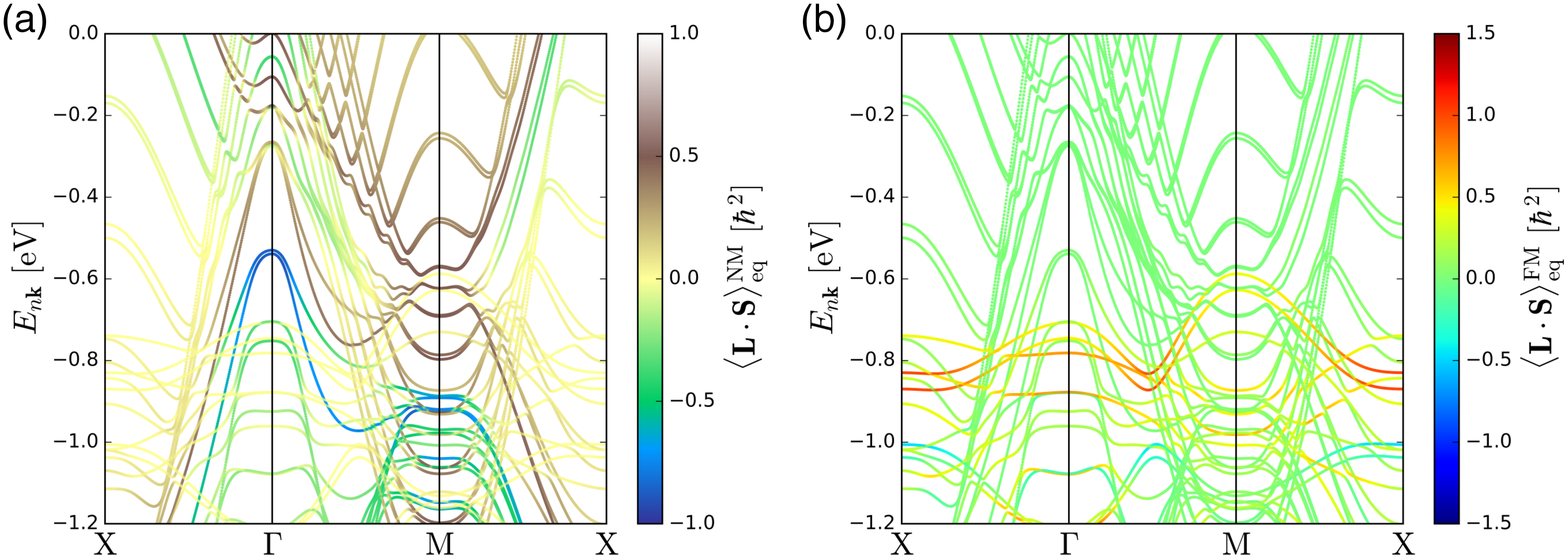}
\caption{
\label{fig:band_soc_NM_FM_onsite}
Band structure and the spin-orbit correlations in the (a) NM and (b) FM regions, when the SOC included in both regions and the $d$ orbitals onsite energies in the FM are lowered by $0.7\ \mathrm{eV}$. For this calculation, we assumed $N_\mathrm{NM}=8$ and $N_\mathrm{FM}=2$.  
}
\end{figure}
%========================================================

%========================================================
\begin{figure}[b!]
\includegraphics[angle=0, width=0.5\textwidth]{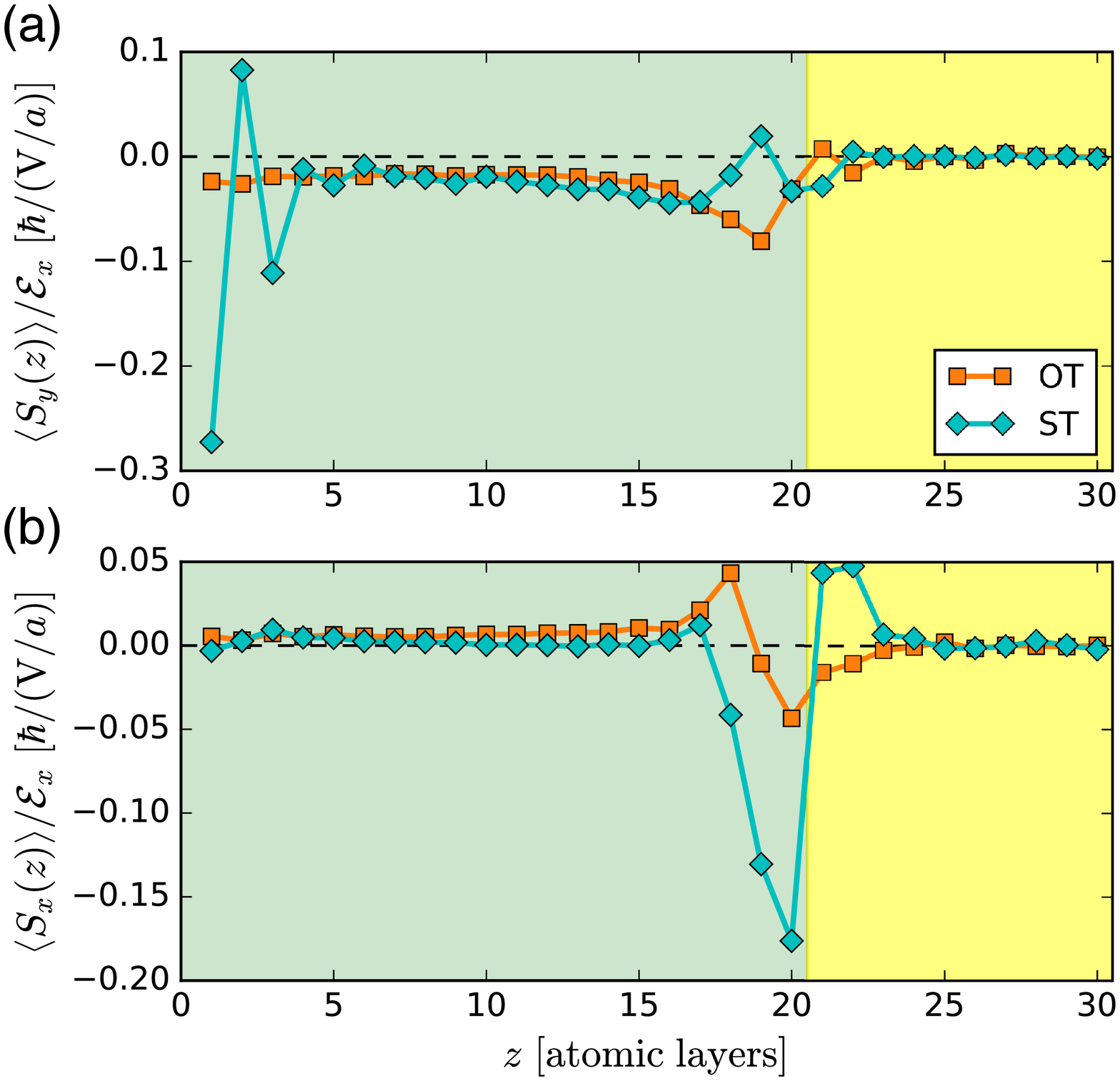}
\caption{
\label{fig:OT_ST_onsite}
Spatial profiles of (a) $\left\langle S_y \right\rangle/\mathcal{E}_x$ and (b) $\left\langle S_x \right\rangle/\mathcal{E}_x$ in the NM ($1\leq z \leq 20$) and FM ($21\leq z \leq 30$) regions. The OT and ST contributions are represented by orange squares and cyan diamonds, respectively. The SOC is included in the NM by Eq.~\eqref{eq:SOC_NM} and the onsite energies of the $d$ orbitals in the FM is shifted by Eq.~\eqref{eq:param_FM_onsite}. The corresponding band structure is shown in Fig.~\ref{fig:band_soc_NM_FM_onsite}. The Fermi energy is set equal to $E_\mathrm{F}=-0.9\ \mathrm{eV}$. 
}
\end{figure}
%========================================================

The result shown in the previous subsection considers the case when the sign of the OT and ST is same. In this subsection, we present a result in another regime where the OT and ST have the opposite signs. To achieve this, we shift the onsite energy of $d$ orbitals in the FM by setting
\begin{eqnarray}
\label{eq:param_FM_onsite}
E_{d_{xy}} = E_{d_{yz}} = E_{d_{zx}} = E_{d_{x^2-y^2}} = E_{d_{z^2}} = -1.2\ \mathrm{eV},
\end{eqnarray}
which is lower than the original system by $0.7\ \mathrm{eV}$ [Eq. \eqref{eq:param_FM}]. The rest of the parameters are unchanged. The spin-orbit correlations in the NM and FM are shown on top of the band structure in Figs.~\ref{fig:band_soc_NM_FM_onsite}(a) and \ref{fig:band_soc_NM_FM_onsite}(b), respectively. Now near $E_\mathrm{F}=-0.9\ \mathrm{eV}$, $\left\langle \mathbf{L}\cdot\mathbf{S} \right\rangle_\mathrm{eq}^\mathrm{NM}<0$ but $\left\langle \mathbf{L}\cdot\mathbf{S} \right\rangle_\mathrm{eq}^\mathrm{FM}>0$, thus we expect the opposite signs for the OT and ST. From the same method [Eq.~\eqref{eq:OT_ST}], we calculate the electric field responses of the OT and ST contributions. Figures~\ref{fig:OT_ST_onsite}(a) and \ref{fig:OT_ST_onsite}(b) show the results for $\left\langle S_y (z) \right\rangle/\mathcal{E}_x$ and $\left\langle S_x (z) \right\rangle/\mathcal{E}_x$, respectively. We find the signs of the OT and ST contributions are opposite in both cases. The result for $\left\langle S_x (z) \right\rangle$ [Fig.~\ref{fig:OT_ST_onsite}(b)], which is the interband contribution from the Fermi sea, is understood as the following. Although there are many FM bands with  $\left\langle \mathbf{L}\cdot\mathbf{S} \right\rangle_\mathrm{eq}^\mathrm{FM}<0$ below $E_\mathrm{F}=-0.9\ \mathrm{eV}$, the hotspots for the OHE and SHE in the NM is concentrated in the energy range $-1.0\ \mathrm{eV} \lesssim E_{n\mathbf{k}} \lesssim 0.0\ \mathrm{eV}$ \cite{Go2018_supp} thus the torque contribution from the FM bands with $\left\langle \mathbf{L}\cdot\mathbf{S} \right\rangle_\mathrm{eq}^\mathrm{FM}<0$ is negligible and the major contribution is from the states in an energy range $-1.0\ \mathrm{eV} \lesssim E_{n\mathbf{k}} \lesssim -0.9\ \mathrm{eV}$.

\subsection{F. \ Role of the Interface Hoppings for the Orbital Torque}

%========================================================
\begin{figure}[t!]
\includegraphics[angle=0, width=0.5\textwidth]{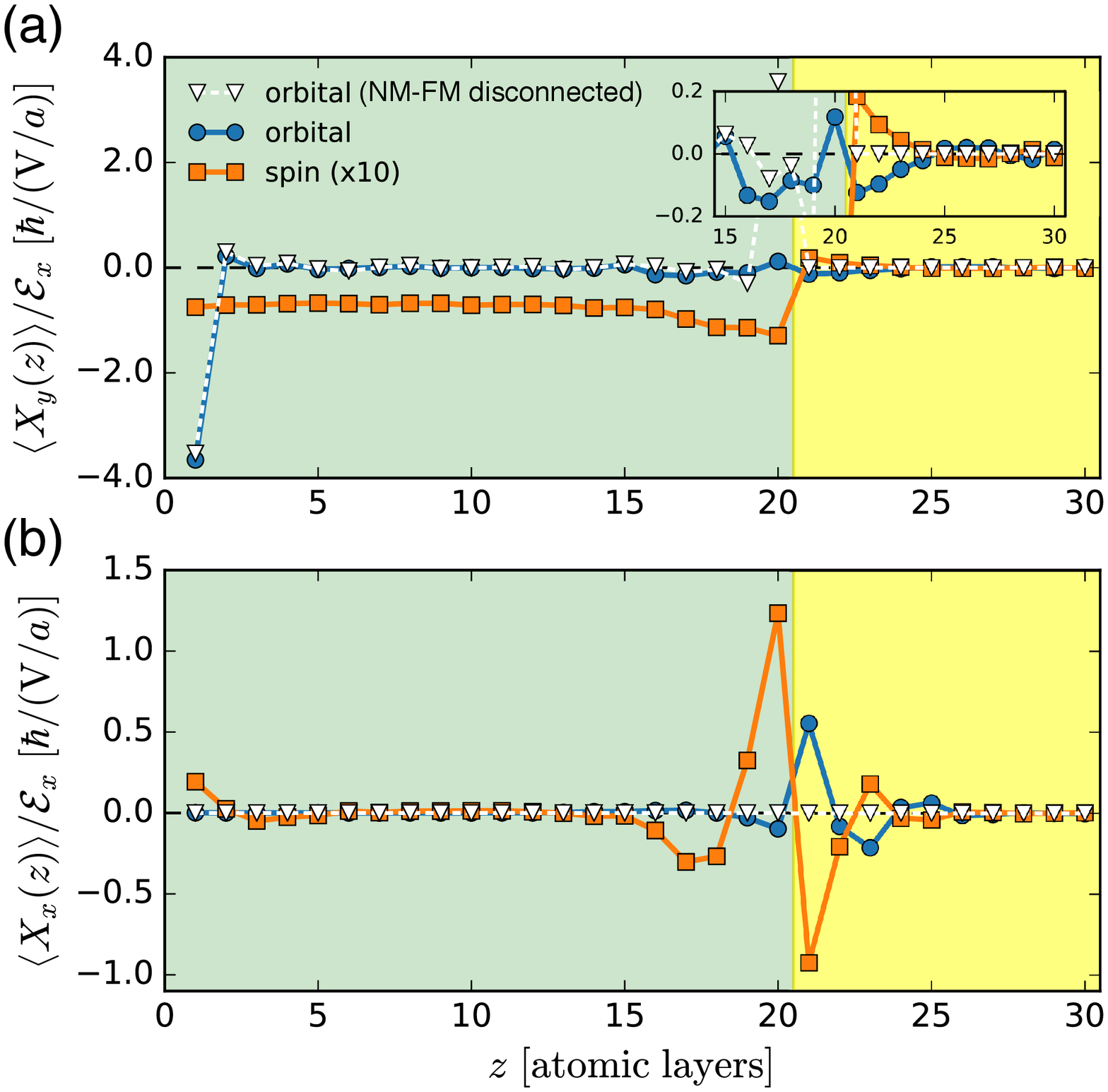}
\caption{
\label{fig:spatial_profile_interface}
(a) $\langle X_y(z) \rangle / \mathcal{E}_x$  and (b) $\langle X_x(z) \rangle / \mathcal{E}_x$ as a function of $z$ for $E_{\rm F}=-0.9\ \mathrm{eV}$, when the relative sign of the interface hoppings is flipped [Eq.~\eqref{eq:param_interface_flip}]. Blue circles (orange squares) depict the orbital (spin multiplied by factor 10) accumulation profile in the NM ($1\le z \le 20$) and the FM ($21 \le z \le 30$). White inverted triangles depict the orbital accumulation profile when the NM and FM are disconnected. The inset in (a) presents a magnified plot near the NM/FM interface.
}
\end{figure}
%========================================================

In this section, we demonstrate crucial role of the interface hoppings for the OT. For the orbital injection from the NM to the FM in the tight-binding model presented in the Letter, the orbital information in the $p$ orbitals in the NM should be transferred to the $d$ orbitals in the FM. At the interface, two types of hoppings are crucial for this:
\begin{subequations}
\label{eq:interface_hoppings}
\begin{eqnarray}
t_{pd}^\sigma &=&
\bra {p_z^\mathrm{NM}} H_\mathrm{hop} \ket{d_{z^2-x^2}^\mathrm{FM}},
\\
t_{pd}^\pi &=&
\bra {p_x^\mathrm{NM}} H_\mathrm{hop} \ket{d_{zx}^\mathrm{FM}}.
\end{eqnarray}
\end{subequations}
Once a state carrying finite OAM, $\ket{L_y^{(p)}=\pm1}=\ket{p_z}\pm i\ket{p_x}$ for example, is induced in the NM, the interface hoppings in Eq.~\eqref{eq:interface_hoppings} can generate a state $\ket{L_y^{(d)}=\pm 2}=\ket{d_{z^2-x^2}^\mathrm{FM}} \pm i \ket{d_{zx}^\mathrm{FM}}$ that also carries net OAM.

Thus, the relative sign of $\gamma_{pd\sigma}$ and $\gamma_{pd\pi}$ is crucial, by which the OT changes the sign. In the tight-binding model used in the Letter, we assume the same sign for $\gamma_{pd\sigma}$ and $\gamma_{pd\pi}$ [Eq.~\eqref{eq:param_interface}]. In order to demonstrate this effect, we present calculation results for $\left\langle X_y (z) \right\rangle/\mathcal{E}_x$ and $\left\langle X_x (z) \right\rangle/\mathcal{E}_x$ in Figs.~\ref{fig:spatial_profile_interface}(a) and \ref{fig:spatial_profile_interface}(b), respectively, by assuming 
\begin{eqnarray}
\label{eq:param_interface_flip}
\gamma_{pd\sigma}=-0.4, \ \ \gamma_{pd\pi}=0.1,
\end{eqnarray}
which is to be compared with Fig.~3 of the Letter. We find that $\left\langle L_y (z) \right\rangle/\mathcal{E}_x$ is unchanged near $z=1$, which is away from the interface. However, near the interface ($z=20$) and in the FM region ($21 \leq z \leq 30$), the sign of the $\left\langle L_y (z) \right\rangle/\mathcal{E}_x$ in Fig.~\ref{fig:spatial_profile_interface}(a) is opposite to that in Fig.~3(a) of the Letter. As a consequence, $\left\langle S_y (z) \right\rangle/\mathcal{E}_x$, which is converted from the injected orbital angular momentum, also changes the sign [Fig.~\ref{fig:spatial_profile_interface}(a)]. Since $\left\langle S_x (z) \right\rangle/\mathcal{E}_x$ precesses along the magnetization by the exchange interaction and $\left\langle L_x (z) \right\rangle/\mathcal{E}_x$ follows by the SOC in the FM, the signs of $\left\langle L_x (z) \right\rangle/\mathcal{E}_x$ and $\left\langle S_x (z) \right\rangle/\mathcal{E}_x$ in Fig.~\ref{fig:spatial_profile_interface}(b) are flipped compared to Fig.~3(b) of the Letter. 

Therefore, the interface crystallinity is crucial for the generation of the OT. In dirty interface, the interface hoppings, such as Eq.~\eqref{eq:interface_hoppings}, are randomized, and this reduces the magnitude of the OT. On the other hand, the spin injection is not affected by the relative sign of the interface hoppings, thus the ST is less susceptible to the interface crystallinity.

%\bibliography{bib_orbital_torque_supp}

\end{document}